\documentclass[11pt,a4paper]{article}
\pdfoutput=1
\usepackage{jheppub}
\usepackage[T1]{fontenc}
\usepackage{booktabs}
\usepackage{xcolor}
\usepackage{bbm}
\usepackage{orcidlink}

\let\fullsizetableofcontents\tableofcontents
\renewcommand{\tableofcontents}{\begingroup\footnotesize\fullsizetableofcontents\endgroup}
\makeatletter\gdef\@fpheader{Prepared for submission to JHEP}\makeatother
\graphicspath{{./}}

\newcommand{\R}{\mathbb{R}}
\newcommand{\Z}{\mathbb{Z}}
\newcommand{\CP}{\mathbb{CP}}
\newcommand{\Ric}{\mathrm{Ric}}
\newcommand{\Hess}{\mathrm{Hess}}
\newcommand{\vol}{\mathrm{vol}}
\newcommand{\Tone}{T^{1,1}}
\newcommand{\Ypq}{Y^{p,q}}
\newcommand{\dP}[1]{\mathrm{dP}_{#1}}
\newcommand{\Gcan}{G_{\rm can}}

\title{Numerical Sasaki--Einstein metrics and harmonic forms on del Pezzo links}

\author[a]{Nakwoo Kim\,\orcidlink{0000-0002-4182-8459}}
\author[b]{Sejin Kim\,\orcidlink{0000-0003-1176-4199}}
\author[a]{and Hoseob Shin\,\orcidlink{0009-0005-5662-2399}}

\affiliation[a]{Department of Physics and Research Institute of Basic Science,
Kyung Hee University, Seoul 02447, Republic of Korea}
\affiliation[b]{Center for AI and Natural Sciences,
Korea Institute for Advanced Study, Seoul 02455, Republic of Korea}

\emailAdd{nkim@khu.ac.kr}
\emailAdd{sejin@kias.re.kr}
\emailAdd{hoseob0610@khu.ac.kr}

\abstract{We numerically construct the Sasaki--Einstein metric on the link of the
cone over the second del Pezzo surface $\dP2$ and two primitive harmonic basic
$(1,1)$-forms at its irregular volume-minimizing Reeb vector. Being toric, 
the metric in symplectic coordinates is encoded in a single convex function 
on a polygon. We approximate the correction to the canonical Guillemin potential 
in two ways: polynomial expansion and neural networks. 
The polynomial fit achieves a held-out mean-squared Monge--Amp\`ere
residual below $10^{-13}$, in contrast to the $10^{-2}$ plateau
for the non-volume-minimizing regular Reeb vector.
We validate our method against closed-form metrics of $Y^{p,q}$ using curvature invariants, and also against the numerical result of Doran et al. (2007) for the K\"ahler--Einstein metric of $\dP3$ using the 
Laplacian spectrum of low torus-invariant modes. The integrated
Riemann-squared curvature of the $\dP2$ link agrees to two parts in $10^{9}$
with the closed form conjectured from the dual field theory. Our data for the metric and harmonic forms can be used to study warped non-conformal holographic IIB backgrounds, the analogues
of the Klebanov--Tseytlin solution on the conifold.}


\arxivnumber{2609.25857}

\begin{document}
\maketitle

\section{Introduction}
\label{sec:intro}

A Sasaki--Einstein metric is an Einstein metric on the compact link of a
Ricci-flat K\"ahler cone. Through the AdS/CFT correspondence
\cite{Maldacena:1997re} an $AdS_5\times{\rm SE}_5$ background of type IIB is
dual to a four-dimensional ${\cal N}=1$ superconformal field theory
\cite{Kehagias:1998gn,Acharya:1998db,Morrison:1998cs}. In five dimensions the
explicit toric examples are, up to finite quotients, the round $S^5$, the
homogeneous $\Tone$, whose cone is the conifold \cite{Candelas:1989js}, and the
cohomogeneity-two family $L^{a,b,c}$
\cite{Cvetic:2005ft,Cvetic:2005vk,Martelli:2005wy}, which contains the
cohomogeneity-one $\Ypq$ \cite{Gauntlett:2004yd,Gauntlett:2004hh} as
$L^{p-q,p+q,p}$. The field theory duals are known and have been checked in
detail in each case
\cite{Klebanov:1998hh,Benvenuti:2004dy,Butti:2005sw,
Franco:2005sm,Benvenuti:2005ja}.

Several quantities relevant to the dual field theory can be computed without
an explicit metric. The Reeb vector, the volume of the link and the volumes of
supersymmetric cycles follow from the toric data by volume minimization
\cite{Martelli:2005tp,Martelli:2006yb}, in agreement with field-theoretic
$a$-maximization \cite{Intriligator:2003jj,Butti:2005vn}. We refer to these
quantities as protected. Other observables, including the masses of long
Kaluza--Klein multiplets, depend on the metric. Numerical metrics provide
access to this spectrum when no closed-form expression is available.

The del Pezzo cones are the natural place to begin. The quiver gauge theories
on D3-branes at these singularities were among the first toric examples to be
worked out \cite{Feng:2000mi,Feng:2001xr,Feng:2001bn,Beasley:2001zp,Franco:2005rj},
and the $\dP2$ and $\dP3$ theories each come in several Seiberg-dual phases. 
Closed forms for the metrics are known for
the first two: the cone over $\dP0$ is $\mathbb C^3/\Z_3$, and the link of the
cone over $\dP1$ is $Y^{2,1}$. For $\dP2$ and $\dP3$ they are not, and this
paper computes the Ricci-flat K\"ahler metrics on both cones numerically. The
two are different in kind. On the cone over $\dP3$ the Reeb vector is \emph{regular},
so the transverse metric is the K\"ahler--Einstein metric of the surface
itself, which exists \cite{TianYau:1987,Siu:1988,Tian:1990} and has been
computed numerically before \cite{Doran:2007zn,BunchDonaldson:2008}; we
reproduce that computation at higher precision and check ours against it. On
the cone over $\dP2$ there is no smooth quotient surface: $\dP2$ carries no
K\"ahler--Einstein metric \cite{Tian:1990}, its volume-minimizing Reeb vector
is \emph{irregular}, and the Sasaki--Einstein metric on the link, which exists
\cite{Futaki:2006cc}, has not to our knowledge been constructed. That metric,
and the two primitive harmonic basic forms on it in section~\ref{sec:oneone},
are the main contributions of this work. Table~\ref{tab:targets} collects the three
problems.

\begin{table}[htbp]
\centering
\footnotesize
\begin{tabular}{llll}
\toprule
problem & Reeb vector & existence & held-out residual\\
\midrule
K\"ahler--Einstein on $\dP3$ & regular &
exists \cite{TianYau:1987,Siu:1988,Tian:1990} & $9.1\times10^{-14}$\\
Sasaki--Einstein on the $\dP2$ link & $b^\ast$, irregular & exists
\cite{Futaki:2006cc} & $6\times10^{-14}$\\
K\"ahler--Einstein on $\dP2$ & regular & does not exist
\cite{Matsushima:1957,Tian:1990,WangZhu:2004,Mabuchi:1987} & floor
$\approx4\times10^{-2}$\\
\bottomrule
\end{tabular}
\caption{The three del Pezzo problems. The third column records what is known
about existence, from theorems; an external numerical reference exists only for
$\dP3$ \cite{Doran:2007zn}, and for $\dP2$ a closed form for one integrated
curvature invariant (section~\ref{sec:dp2se}). At a regular Reeb vector the transverse
geometry is a surface and the transverse equation is the K\"ahler--Einstein
equation on it; at an irregular one the same equation holds on the polygon, but
the transverse geometry it describes exists only locally,
while the five-dimensional link is globally smooth. The residual is evaluated on sample points
withheld from the fit (section~\ref{sec:param}); a
floor denotes a residual plateau over the tested range of approximation orders.}
\label{tab:targets}
\end{table}

The examples are organized by the available reference. We begin with $\Tone$,
whose canonical potential solves the equation exactly, and then study $\Ypq$,
whose metrics are known in closed form from Gauntlett, Martelli, Sparks and
Waldram (GMSW) \cite{Gauntlett:2004yd,Gauntlett:2004hh}. The third reference is
the numerical $\dP3$ metric of Doran, Headrick, Herzog, Kantor and Wiseman
(DHHKW) \cite{Doran:2007zn}, the positive control. Finally, we examine the
$\dP2$ cone at its regular Reeb vector, where a non-existence theorem applies
\cite{WangZhu:2004,Mabuchi:1987} and the negative control records the method's
behaviour without identifying the cause of numerical failure, and at its
irregular minimizer, where a Sasaki--Einstein metric exists and no independent
metric is available for comparison.

We use the symplectic formulation of
\cite{Martelli:2005tp,Doran:2007zn}, in which the toric cone metric is specified
by a convex potential on a polygon and Ricci-flatness becomes a scalar
Monge--Amp\`ere equation. The same equation applies at irregular Reeb vectors.
We represent the smooth correction in two ways: a \emph{polynomial} expansion in the
slice coordinates, symmetrized when indicated, and a multilayer perceptron
with the same two inputs. The polynomial fits achieve smaller residuals on
the examples studied and supply the most precise metric approximations.
The \emph{networks} provide a second function class without a polygon-adapted basis.
They allow us to test recovery of an unimposed symmetry and to compare
residual plateaus across parametrizations. Both classes use the same equation,
domain and canonical potential, so this comparison does not establish
independence from the full computational pipeline.

Related numerical methods address several geometric settings. DHHKW reduced the toric problem to an equation for
a convex function on the polygon and solved it on $\dP3$ three ways, one of
them in the ansatz used here, a Guillemin potential corrected by
$D_6$-invariant polynomials. Both the reduction and that ansatz are theirs;
we change the coefficient-fitting procedure and extend the validation and target set. Bunch and Donaldson
\cite{BunchDonaldson:2008} computed extremal metrics on toric surfaces from a
fourth-order equation on the polygon, among them the hexagon, where the
extremal metric is by their own account the K\"ahler--Einstein metric of
$\dP3$. Headrick and Wiseman \cite{HeadrickWiseman:2007sol} are the nearest
precedent to our pentagon. A toric Fano surface that admits no
K\"ahler--Einstein metric still admits a K\"ahler--Ricci soliton
\cite{WangZhu:2004}, and it is that soliton on the compact $\dP2$ that they
compute: a different equation on a different space, unobstructed by the
criterion that obstructs ours. The scheme of Berman et al.\
\cite{Berman:2020jas} is aimed at our equation but has, as far as we have
found, not been implemented. Neural and spectral methods for
compact Calabi--Yau manifolds
\cite{Ashmore:2019wzb,Douglas:2020hpv,Larfors:2022nep,Butbaia:2024xgj} solve
for Ricci-flat K\"ahler metrics on compact complex manifolds with a global
holomorphic volume form, and treat neither cones, nor links, nor irregular
foliations. Neural solvers for the Einstein condition on other spaces
\cite{Hirst:2025seh,SchettiniGherardini:2026bdb,DeLuca:2024njg} and for
$G_2$-structures on contact Calabi--Yau seven-manifolds \cite{Heyes:2026rch}
use parametrizations different from ours, and no Reeb vector appears in
them;
a proceedings overview of physics-informed networks in differential geometry
is \cite{Hirst:2026cpm}.

The residual plateaus on $\dP2$ are interpreted in light of the known
obstruction, not as numerical proofs of non-existence. The $\dP3$ comparisons
validate the implementation and compare finite approximations; they do not
establish superiority of one parametrization at matched degree and cost.
Protected quantities check the toric data and normalization rather than the
metric approximation. Our main results are the numerical metric and basic
forms at the irregular Reeb vector of $\dP2$. The $\dP3$ forms were already
computed by DHHKW; here we use a different parametrization and construct a
period-fixed basis. Kaluza--Klein spectroscopy of the new $\dP2$ metric is
reported separately.

Section~\ref{sec:setup} defines the geometry, equation, parametrizations and
validation quantities. Section~\ref{sec:closedform} treats $\Tone$ and $\Ypq$,
including the conditioning of the polynomial basis.
Section~\ref{sec:dp3} compares with DHHKW and examines symmetry recovery.
Section~\ref{sec:dp2} presents the $\dP2$ metric and the obstructed
control. Section~\ref{sec:oneone} constructs the primitive
harmonic forms, and section~\ref{sec:discussion} summarizes the results and
limitations. Appendices~\ref{app:potential}--\ref{app:admissibility} give the
potential identities, numerical protocols, the $\Ypq$
reference metric, the conventions of the comparison with DHHKW, and the
admissibility and completeness of the harmonic-form trial space.\footnote{Code, random seeds and stored minimizers:
\url{https://github.com/nkim-khsb/toric-se-ml}.}

\emph{Note added.} After this paper was submitted, Buttsworth, Hadden, Heyes,
Platt and Wiseman \cite{Buttsworth:2026nap} computed the K\"ahler--Einstein
potential on $\dP3$ from the same Monge--Amp\`ere equation, with the correction
to the canonical potential expanded in Chebyshev polynomials on a fundamental
domain of the hexagon and solved by Newton's method in high-precision
arithmetic, reaching a residual of order $10^{-27}$ certified by 
interval arithmetic. We have checked that our degree-$18$
metric of section~\ref{sec:dp3} agrees with theirs to one part in $10^{8}$
on average and to one part in $10^{7}$ at worst, near a vertex. Our result
for the first $T^2\rtimes D_6$-invariant eigenvalue, $6.32277$ in table~\ref{tab:fullsector}, is within their certified interval $[6.2367,\,6.5976]$. Their construction
relies on the compact surface and does not extend to the irregular Reeb vector
of $\dP2$. The comparison is in the repository of footnote~1.

\section{The geometry, the equation, and the checks}
\label{sec:setup}

We formulate the metric problem on the moment polygon and describe the
numerical method and the checks used to assess its solutions. Further background and derivations are given in
\cite{Martelli:2005tp,Doran:2007zn}.

\subsection{From the cone to the polygon: the Monge--Amp\`ere equation and the Reeb vector}
\label{sec:setupgeom}

Let $Y$ be a compact five-dimensional manifold and $C(Y)=\R_+\times Y$ the
real six-dimensional metric cone over it, with
$g_C=dr^2+r^2g_Y$, and $\Ric(g_C)=0$ holds exactly when $\Ric(g_Y)=4g_Y$, so
that $g_Y$ has scalar curvature $20$. The two Einstein conditions are therefore equivalent. The
Sasakian structure is carried by a single vector field, the Reeb vector $\xi=J(r\partial_r)$. Then the link metric splits as
$g_Y=g_T+\eta\otimes\eta$ with $\eta(\xi)=1$, and the Einstein condition
becomes $\Ric(g_T)=6g_T$ on the transverse part $g_T$, whose scalar curvature
is then $24$.

The Reeb flow is regular when all orbits close and the circle action is free,
quasi-regular when all orbits close but finite stabilizers occur, and irregular
otherwise. In the irregular case a generic orbit is dense in a subtorus of the
$T^3$, of dimension two or three (two on the $\dP2$ cone below, where $b^{\ast}$
has a vanishing component). In the irregular case the leaf space is neither a
manifold nor an orbifold, and $g_T$ is defined only transversally. The cone and the link, on the other hand,
exist in all three cases. That is why a toric Sasaki--Einstein link can sit over a
polygon whose surface carries no K\"ahler--Einstein metric.

\begin{figure}[t]
\centering
\includegraphics[width=0.94\textwidth]{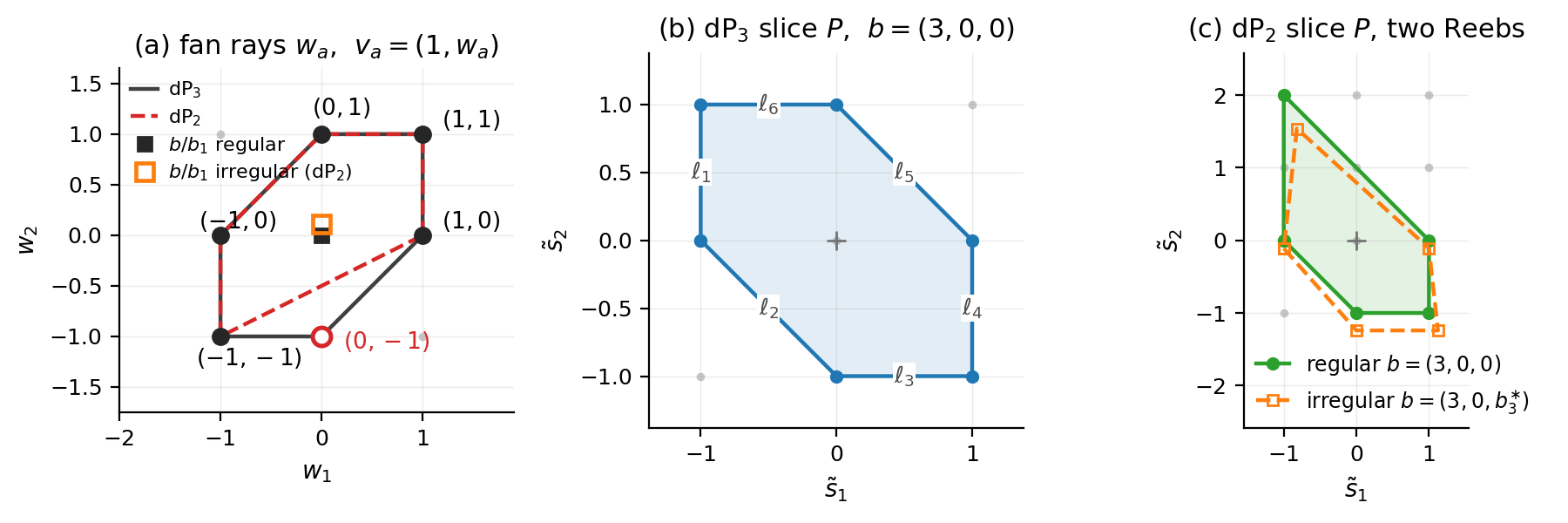}
\caption{The toric data of this paper, drawn in the display chart $\tilde s=3s$. Left
to right, the fan rays $w_a$ fix the inward normals $v_a=(1,w_a)$, and the
slice polygon $P$ is the section of the moment cone at $\langle b,t\rangle=1$.
The marked interior point is $b/b_1$, whose rationality is the arithmetic that
separates closed Reeb orbits from dense ones.}
\label{fig:polytopes}
\end{figure}

When $C$, or equivalently $Y$, is toric, one can read the same objects off the toric diagram. Being toric means that a three-torus
acts on the cone by isometries and also preserves the complex structure $J$. The moment
cone is then read as a set of inequalities,
$\mathcal C=\{y\in\R^3\mid\ell_a(y)\equiv\langle v_a,y\rangle\ge0\}$, with inward
primitive normals $v_a$ and one facet function $\ell_a(y)$ for each facet. On a
toric Calabi--Yau cone a lattice basis exists in which every normal reads
$v_a=(1,w_a)$ \cite{Martelli:2005tp}. The two-dimensional picture of the $w_a$ is called the (fan) toric diagram, and their data for $\dP2, \dP3$ are given in figure~\ref{fig:polytopes}.

In action--angle coordinates the metric takes a block form found by Abreu
\cite{Abreu:1998xhs,Abreu:2000xhs},
\begin{equation}
g=G_{IJ}\,dy^I\,dy^J+G^{IJ}\,d\phi_I\,d\phi_J,
\qquad G_{IJ}=\frac{\partial^2G}{\partial y^I\,\partial y^J},
\label{eq:block}
\end{equation}
where $G^{IJ}$ is the inverse Hessian and $I,J=1,2,3$. The three angles $\phi_I$ parametrize a three-torus, and being toric means $G$ is independent of $\phi_I$. The whole six-dimensional metric is therefore determined by a single convex function $G$, which is called the symplectic potential. The boundary behaviour of $G$ is fixed by smoothness of the metric: as $\ell_a\rightarrow 0$, $G$ must have the Guillemin form
$\tfrac12\ell_a\ln\ell_a$ \cite{Guillemin:1994kah,Abreu:2000xhs}. For $\CP^2$, the Guillemin potential itself gives the Fubini--Study metric.

In the coordinate basis given above, the Reeb vector is $b$, with its 
associated Killing vector field $\xi=\sum_Ib_I\partial_{\phi_I}$. The facet
function of the Reeb vector is written $\ell_b(y)\equiv\langle b,y\rangle$, and $b/b_1$ is an
interior point of the toric diagram. The orbit type is then an arithmetic
question on $b$. The orbits close when $b$ is proportional to a lattice vector
and do not close when it is not, so a rational $b/b_1$ means regular or
quasi-regular and an irrational one means irregular. Which of the two closed
cases occurs is a question about isotropy and not about rationality: writing
$\hat b$ for the primitive lattice vector along $b$, the circle action is free,
and the leaf space a manifold, exactly when $|\det[\hat b,v_a,v_{a+1}]|=1$ on
every pair of adjacent facets. That determinant is $1$ for $\Tone$ and for
$\dP3$, whose $b/b_1$ are $(1,\tfrac12,\tfrac12)$ and $(1,0,0)$, and it is not
for $Y^{7,3}$, whose $b/b_1$ is rational but whose leaf space is an orbifold.

Ricci-flatness of the cone is now a single scalar Monge--Amp\`ere equation
\cite{Martelli:2005tp},
\begin{equation}
\ln\det\Hess G+2\,\partial_{y^1}G+c=0.
\label{eq:coneMA}
\end{equation}
The direction $\partial_{y^1}$ is fixed
by the gauge choice of our fan $v_a=(1,w_a)$ and $c$ is an integration constant.
Two forms of the equation appear below, and \eqref{eq:coneMA} --- the cone
form --- is the one the optimizer minimizes at every target, the entry gate of
section~\ref{sec:t11} excepted, which is posed in the transverse form. At a regular Reeb
vector one may instead impose the transverse K\"ahler--Einstein condition on the
quotient surface, where it reads
$\ln\det\Hess G+2\Lambda G-\nabla G\cdot(2\Lambda \tilde s -\gamma)-c=0$ with
$\Lambda$ the transverse Einstein constant, $\gamma$ an affine datum fixed by
the polytope, and $\tilde s$ a coordinate on that polytope
\cite[(3.25)]{Doran:2007zn}. We use that transverse form twice:
to calibrate on $\Tone$ in section~\ref{sec:t11}, and as the normalization
bridge to DHHKW in section~\ref{sec:dp3grading}. The transverse equation itself
exists at every Reeb vector: restricted to the slice, the cone form is exactly
this equation with $\Lambda=3$ and a $\gamma$ fixed by $b$, whether or not $b$
is rational (appendix~\ref{app:potential}, eq.~\eqref{eq:slicema}). What an
irregular $b$ removes is its interpretation as the K\"ahler--Einstein equation of
a quotient surface, and that interpretation is what both uses above rely on. The cone
form, which needs no quotient, carries the paper.
The link sits at a definite slice of the cone. The normalization $|\xi|=1$
gives $r^2=2\ell_b(y)$, so the link is the surface $\ell_b(y)=\tfrac12$. For a Sasaki--Einstein metric the Reeb vector is not arbitrary:
it is the unique minimizer $b^{\ast}$ of the volume functional, which is a
strictly convex rational function of $b$ chosen from the interior of the dual cone
$\mathcal C^{\vee}=\{b\mid\langle b,y\rangle\ge0\ \text{for all}\ y\in\mathcal C\}$.
The constraint $b_1=3$ below is not a choice: in the gauge $v_a=(1,w_a)$ the
holomorphic volume form has charge $b_1$ under the Reeb vector, and the
Calabi--Yau condition in complex dimension three fixes that charge to three \cite{Martelli:2005tp}.
\begin{equation}
b^{\ast}=\operatorname*{arg\,min}_{b_1=3,\;b\in\mathrm{int}\,\mathcal C^{\vee}}\vol(Y_b),
\qquad
\frac{\vol(Y_b)}{\pi^3}
=\frac1{b_1}\sum_{a}
\frac{\bigl|\det[v_{a-1},v_a,v_{a+1}]\bigr|}
{\bigl|\det[b,v_{a-1},v_a]\,\det[b,v_a,v_{a+1}]\bigr|}.
\label{eq:volmin}
\end{equation}
Minimization returns $(3,0,0)$ on $\dP3$ and
$\bigl(3,0,\tfrac3{16}(19-3\sqrt{33})\bigr)$ on $\dP2$ --- the latter is
eq.~(3.48) of Martelli, Sparks and Yau (MSY) \cite{Martelli:2005tp}, written in a different lattice
basis, and the components of $b$ are basis dependent while its irrationality is
not --- and having an 
irrational entry means that the Sasaki--Einstein structure on the $\dP2$ cone
is irregular. Table~\ref{tab:orbits} lists and classifies the examples we consider in this work. For $\Ypq$, we follow the convention of \cite{Martelli:2005tp} and set $w_1=(0,0),w_2=(p-q-1,p-q),w_3=(p,p),w_4=(1,0)$; for $\Tone=Y^{1,0}$ this is the unit square $(0,0),(0,1),(1,1),(1,0)$.

\begin{table}[htbp]
\centering
\footnotesize
\begin{tabular}{llll}
\toprule
class & generic orbit & leaf space & examples treated here\\
\midrule
regular & closed, free & smooth surface & $\Tone=Y^{1,0}$, $\dP3$ link\\
quasi-regular & closed, finite isotropy & orbifold & $Y^{7,3}$\\
irregular & dense in a subtorus & neither manifold nor orbifold & $Y^{2,1}$, $Y^{3,2}$, $\dP2$ link at $b^{\ast}$\\
\bottomrule
\end{tabular}
\caption{The three types of Reeb orbit, the leaf space in each case, and the targets of this paper that realize them.}
\label{tab:orbits}
\end{table}
\subsection{The ansatz, the objective, and the conventions}
\label{sec:param}

We parametrize the unknown part of $G$ on a real two-dimensional slice. The radial projection is
$t=y/\ell_b(y)$, the slice polygon is $P=\{t\in\mathcal C\mid\langle b,t\rangle=1\}$, and
$s$ is a chart on it, fixed below. Homogeneity
fixes the whole radial dependence, so the general solution reads
$G=\tfrac12\ell_b\ln\ell_b+\ell_bF(t)$ with $F$ arbitrary on the slice. The unknown function is then a
single correction to a canonical potential which is fixed by the lattice data including $b$,
\begin{equation}
G=\Gcan+\ell_b\,\psi(s),
\qquad
\Gcan=\tfrac12\,\ell_b\ln\ell_b
+\tfrac12\,\ell_b\sum_a\langle v_a,t\rangle\ln\langle v_a,t\rangle,
\label{eq:ansatz}
\end{equation}
where $\Gcan$ is the Guillemin form on the slice with factor
$\ell_b$ which is required by homogeneity. This ansatz spans MSY's moduli of
symplectic potentials at every $b$. The difference from any member of that family is $\ell_b$ times
a function smooth on the closed polygon, hence absorbable into $\psi$. Table~\ref{tab:dimensions} summarizes how the spaces of different dimension are related.

\begin{table}[htbp]
\centering
\footnotesize
\begin{tabular}{lll}
\toprule
object & dimension & role\\
\midrule
K\"ahler cone $C(Y)$ & $\dim_{\mathbb C}=3$ & where the equation is posed\\
link $Y$ & $\dim_{\R}=5$ & the Sasaki--Einstein space\\
transverse geometry & $\dim_{\R}=4$ & K\"ahler always, Einstein at a solution; a surface only for regular $b$\\
slice polygon $P$ & $\dim_{\R}=2$ & domain of unknown function $\psi$\\
\bottomrule
\end{tabular}
\caption{The spaces the construction moves between and the role each one plays. The unknown function $\psi$ lives on the last of them.}
\label{tab:dimensions}
\end{table}

Choose affine coordinates $(s_1,s_2)$ on the slice by writing 
\begin{equation}
    t = t_0 + s_1 e_1 + s_2 e_2 ,
\label{eq:slicechart}
\end{equation}
where $t_0=b/|b|^2$ is the point of the slice plane closest to the origin
and $e_1,e_2$
are an orthonormal basis of $\ker b$, the plane
$\langle b,y\rangle=0$; we take the two trailing right-singular vectors of $b$
regarded as a $1\times3$ matrix, and nothing below depends on that choice,
since a rotation of the chart leaves the residual unchanged. Throughout the paper lower-case $i,j,k$
run over the two slice coordinates $s_i$, capital $I,J$ over the three cone
coordinates $y^I$ of section~\ref{sec:setupgeom}, and $a$ labels facets. Then
$\ell_a(t)=\langle v_a,t_0\rangle+s_i\langle v_a,e_i\rangle$, so each
$\ell_a(t)=0$ is a line in the $(s_1,s_2)$ plane, and these lines are the edges
of the slice polygon.
Figure~\ref{fig:polytopes} draws the polygons in the display chart
$\tilde s=3s$, which puts the vertices of the hexagon on the integer lattice, $\ker b$
being a coordinate plane at its $b=(3,0,0)$. At the irregular $b^{\ast}$ of
$\dP2$ no rescaling does that, and the
dashed pentagon there is drawn in the same chart for comparison.

The residual operator is
\begin{equation}
\mathcal R[\psi]=\ln\bigl|\det\Hess G\bigr|+2\,\partial_{y^1}G,
\label{eq:resop}
\end{equation}
up to its free constant, and $\psi$ solves the problem exactly when
$\mathcal R$ is constant on the slice. We minimize the mean of
$(\mathcal R+c)^2$ over the samples, fitting $c$ together with the
coefficients. The ansatz carries the homogeneity and the facet asymptotics, and
there are no penalties; positive definiteness of $\Hess G$, which $\ln|\det|$
does not see, is checked separately below, and a zero of the sampled objective
is a solution of the equation at the samples. Unless a different statistic is explicitly specified, a reported
Monge--Amp\`ere residual is this mean square: a residual of $10^{-14}$ means that $\mathcal R$ departs
from constancy by about $10^{-7}$, and that is the scale to keep in mind when a
residual is set beside a quantity that is not squared, such as a curvature
deviation or a percentage. Other quantities are quoted as stated where they
appear: curvature deviations as relative errors, the harmonic-form energies of
section~\ref{sec:oneone} as a ratio of squared $L^2$ norms.

The objective estimates the polygon average of $(\mathcal R+c)^2$;
its sampling variability is measured in appendix~\ref{app:robust}.
The residual operator is exactly invariant under a change of lattice basis
that preserves the Calabi--Yau gauge $v_a=(1,w_a)$, which is unimodular and
fixes $\partial_{y^1}$, and under radial rescaling, since with $b_1=3$ the
shifts of its two terms cancel (appendix~\ref{app:potential}); our
implementation respects both to machine precision on the targets below. A
residual is therefore comparable across our targets, and between the two
function classes, without a normalization attached to it, with two
qualifications: the sampled mean also depends on the boundary margin
specified below, which is set in one chart and varied in
appendix~\ref{app:robust}, and equal residuals on different targets are equal
errors in the equation, not equal errors in the metric.
In the symplectic coordinates $y$ the domain of $\psi$ is compact and covered
by a single global chart, and the procedure is the same whether $b$ is regular
or irregular.

We first fix the affine freedom in the correction. The shift
$\psi\mapsto\psi+\alpha+\beta\cdot s$ changes only the fitted constant $c$,
so we remove these three flat directions by subtracting an affine function
fitted at three fixed sample points. The polynomial parametrization uses
monomials in $s$ of total degree $2$ through $D$, orthonormalized on the sample
as described in section~\ref{sec:whiten}. Across the targets below,
$D=6$--$18$ gives $14$--$150$ coefficients, with the smallest counts coming
from symmetrized bases. The network is a multilayer perceptron of the same two
coordinates, with two hidden layers of width $w$, $\tanh$ activations and
small initial weights so that $\psi\approx0$. Widths $w=8$--$32$ give
$106$--$1186$ parameters, counting the $w^2+5w+1$ weights and the fitted
constant $c$; the polynomial counts exclude $c$.

Training samples are drawn uniformly on $P$ with a chart distance of at least
$2\times10^{-3}$ from the boundary, where the Hessian of $\Gcan$ diverges.
Held-out points are drawn from the same distribution and never used in
optimization. The main fits use L-BFGS,\footnote{L-BFGS is the limited-memory
Broyden--Fletcher--Goldfarb--Shanno quasi-Newton method
\cite{Liu:1989lbfgs}. It approximates the inverse Hessian using recent
gradient differences and uses a line search. Adam \cite{Kingma:2014adam}
uses running first and second moments of the gradients with a prescribed step
size.} with Adam-only runs identified separately. Table~\ref{tab:cost} in
appendix~\ref{app:robust} lists the model sizes, sample counts and iteration
caps of the principal runs. Unless stated otherwise, all reported residuals
are held-out values. Comparing them with the in-sample residuals tests whether
the fit generalizes beyond its samples; a growing separation indicates
overfitting. Convexity is checked separately, since the objective uses
$\ln|\det|$ and does not enforce positive definiteness.

We then increase the polynomial degree or network width to assess convergence
with the approximation space; the behaviours observed, on solvable and
obstructed targets, are reported in sections~\ref{sec:whiten}
and~\ref{sec:fingerprint}.

\subsection{Checks beyond the residual}
\label{sec:graders}
A small sampled residual does not by itself establish metric accuracy. We
therefore evaluate quantities that depend on the fitted metric but do not
enter the objective, and compare them with exact or independently computed
references. We use four types of check.

The first is a pair of six-dimensional curvature invariants
\begin{equation}
\Phi_1=R^{\mu_1\mu_2}{}_{\mu_3\mu_4}\,R^{\mu_3\mu_4}{}_{\mu_1\mu_2},
\qquad
\Phi_2=R^{\mu_1\mu_2}{}_{\mu_3\mu_4}\,R^{\mu_3\mu_4}{}_{\mu_5\mu_6}\,R^{\mu_5\mu_6}{}_{\mu_1\mu_2},
\label{eq:inv6}
\end{equation}
whose indices run over the six directions of the cone; the former is the usual
Riemann-squared and the latter written alternatively as
$\mathrm{Riem}^{3}$. On a Ricci-flat
six-manifold they are pure Weyl, and neither enters the objective, so
agreeing in them tests the fit against quantities it was not fitted to.

The second is the scalar curvature of the four-dimensional transverse geometry.
With $u^{ij}=(\Hess G_P)^{-1}$, where $G_P=G\big|_{\ell_b=1}$ is the potential
restricted to the slice and $\partial_i\equiv\partial/\partial s_i$, Abreu's formula
\cite{Abreu:1998xhs,Abreu:2000xhs} gives $S=-\partial_i\partial_j u^{ij}$, a
fourth derivative of the potential where the objective is second order. Our
slice $\ell_b=1$ is the double of MSY's characteristic hyperplane
$\ell_b=\tfrac12$. We denote its transverse metric by $\tilde g_T=2g_T$,
where $g_T$ is the link's transverse metric of section~\ref{sec:setupgeom}.
Its Einstein constant is therefore $3$, and the Abreu scalar must come out
\begin{equation}
S\overset{!}{=}12 .
\label{eq:abreu}
\end{equation}

The third is the Laplacian spectrum of the Sasaki--Einstein space, taken on torus-invariant
modes, functions of the moment-map coordinates alone, where it reduces to the
transverse operator $\triangle f=\partial_i(u^{ij}\partial_j f)$, which is
non-positive, the eigenvalue problem being $-\triangle f=\lambda f$. Under a constant rescaling
$g\mapsto cg$ of the transverse metric both $\lambda$ and $S$ scale as $1/c$,
being inverse lengths squared, so spectra are compared only through the
dimensionless ratio $\lambda/S$, in which the factor two above cancels; on the
link at $r=1$, whose transverse metric is half that of our slice, every
eigenvalue is twice the slice value, $\lambda_{\rm link}=2\lambda$.

The fourth is an integrated invariant of the link itself,
$\int_Y|\mathrm{Riem}_Y|^2$, with $\mathrm{Riem}_Y$ the Riemann tensor of the
five-dimensional metric $g_Y$. For a metric cone over an Einstein link it
follows from $\Phi_1$ at $r=1$ by a constant shift, \eqref{eq:riem2cone}
below, and a closed form in the toric data has been proposed for it from the
dual field theory, proved for regular Reeb vectors and conjectural otherwise
\cite{Eager:2010dk,ArabiArdehali:2015iow,Nakayama:2026bls}. Unlike the first
three it is a single number per target, so it tests the metric only in the
average; it is the one check with an external reference that applies at the
irregular Reeb vector.

The four are not evaluated in the same place, nor on the same targets. The
Abreu scalar and the transverse Laplacian, built from the potential, run on
the slice through $G_P$; the curvature invariants, pointwise and integrated,
are evaluated on the link itself, at $r=1$, which by $r^2=2\ell_b$ is the
slice $\ell_b=\tfrac12$. The pointwise invariants require a closed form to
compare against and are used on $\Ypq$ alone; the spectrum requires an
external computation and is used on $\dP3$ alone; the integrated invariant is
used at the volume-minimizing Reeb vector of the pentagon, after calibration on
$\Tone$ and $Y^{2,1}$ (section~\ref{sec:dp2se}). Only the Abreu scalar, whose
required value is fixed by the equation itself, is available everywhere, and at
the volume-minimizing Reeb vector of the pentagon it is the only pointwise
check that applies.

Two quantities a numerical metric is often checked against are deliberately not
on that list: the volumes of the link and of its toric divisors, and the
bottom $E=\lambda_{\mathbf m}(\lambda_{\mathbf m}+4)$,
$\lambda_{\mathbf m}=\langle b,\mathbf m\rangle$, of each charged
Kaluza--Klein tower, the modes dual to the chiral primary mesonic operators
\cite{Martelli:2005tp}. Both follow from the polytope and $b$ alone, so they
test the toric data, Reeb vector and normalization rather than the accuracy
of a metric approximation.

The Abreu scalar is fourth order in the potential. Its constancy over the slice
is the test; its value only fixes the normalization. We track that constancy
through the mean deviation from \eqref{eq:abreu}, which estimates a fixed
quantity and therefore converges. A maximum over finitely many points generally underestimates the supremum
and can depend strongly on grid resolution, particularly near the polygon
boundary. Both are
measured: refining from $n=120$ to $n=360$ across the five $\Ypq$ targets of
section~\ref{sec:sweep} moves the mean in its third digit, and the worst value
by up to a factor two and not always monotonically, settling only on the target
where $\psi$ is smallest. We
therefore quote a worst value only together with the grid it was taken on --- a
$240\times240$ Cartesian grid over the polygon --- and compare targets by the
mean, which is the summary that converges.

The transverse Laplacian needs no separate machinery, but it does need
integrals. Because the block form \eqref{eq:block} has
$\det g=\det G_{IJ}\det G^{IJ}=1$, the operator is in divergence form, so its
eigenvalues are the stationary values of the \emph{Rayleigh quotient}
\begin{equation}
\mathcal{Q}[f]\;=\;\frac{\int_P u^{ij}\partial_if\,\partial_jf}{\int_Pf^2}\;,
\label{eq:rayleigh}
\end{equation}
both integrals taken in the flat measure of the polygon with no boundary
condition to impose. In quantum-mechanical language $\mathcal{Q}[f]$ is the
expectation value of $-\triangle$ in a trial function, and
minimizing it is the Rayleigh--Ritz variational principle for the ground
state. Two of its consequences are used here and again in
section~\ref{sec:oneone}: $\mathcal{Q}$ is invariant under $f\mapsto cf$, so
the normalization of $f$ never has to be fixed, and a finite basis can only
overestimate, so the eigenvalues approach from above those of the metric that is
used, with exact integrals. This variational bound does not control the
error from approximating the metric itself.

We find the spectrum by Rayleigh--Ritz in a polynomial basis: it is that of a
generalized eigenproblem $K_{\rm Lap}\zeta=\lambda M_{\rm Lap}\zeta$ for the coefficient vector
$\zeta$ of $f$ in that basis, whose two matrices are the numerator
and the denominator of \eqref{eq:rayleigh} evaluated on the basis. For a given
metric the accuracy of the spectrum depends on the quadrature and the
trial degree. Section~\ref{sec:dp3grading} reports the values obtained with
the specified finite metric and trial spaces, using deterministic quadrature.

We also evaluate these checks on deliberately incorrect inputs: a wrong
normalization in section~\ref{sec:t11} and the untrained Guillemin metric in
section~\ref{sec:dp3grading}. These comparisons establish their sensitivity
to errors not represented by the training residual alone.

The curvature instrument is calibrated where the answer is known. On the
conifold it returns $\Ric=0$ to $4\times10^{-13}$ pointwise and
$2\times10^{-15}$ in the mean. The quadrature is calibrated the same way, at
the exactly known eigenvalue $\lambda_1=6$ of $\CP^1\times\CP^1$: a
centroid-fan triangulation of the polygon, the centroid joined to each vertex,
with tensor Gauss--Legendre nodes on each triangle reaches
$2.8\times10^{-13}$ with $64$ nodes and the $10^{-14}$ level from a few
hundred on, where Monte-Carlo sampling is still at $2\times10^{-2}$ with
$8\times10^{4}$ points; the deterministic rule is the one we use throughout.
The same run reproduces the two lowest $D_6$-invariant eigenvalues
$6.322773$ and $17.09405$ on $\dP3$, so the calibration and the quoted
spectra come out of one instrument. Those two numbers are in DHHKW's
normalization $S=4$ of \eqref{eq:dhhkwnorm}, not in the $S=12$ that gives
$\lambda_1=6$ above; apart from that calibration value, eigenvalues are quoted
in DHHKW's normalization unless a different scale is explicitly indicated,
and Abreu scalars use our slice normalization.

\section{Closed-form references: \texorpdfstring{$\Tone$}{T(1,1)} and \texorpdfstring{$\Ypq$}{Y(p,q)}}
\label{sec:closedform}

We first test the method on $\Tone$, for which the canonical potential is
exact, and on $\Ypq$, for which a nonzero correction can be compared with the
GMSW closed form. Both polynomial and network parametrizations are used.
The polynomial coefficients enter the potential linearly, although the
Monge--Amp\`ere objective remains nonlinear in them. The network provides a
second approximation class under the specified training protocols.

\subsection{\texorpdfstring{$\Tone$}{T(1,1)}: calibration on an exact solution}
\label{sec:t11}

The first target is $\Tone$. It is the only one whose canonical
potential is already the solution, so it is where the instruments are calibrated and
where the network's entry test is first posed. Volume minimization returns
$b=\bigl(3,\tfrac32,\tfrac32\bigr)$ here. In the chart $s$ the resulting slice
is a tilted quadrilateral not centered on the origin. The offsets follow from
$\ell_a(t_0)=\langle v_a,b\rangle/|b|^2$ with the normals
$v_a=(1,0,0),(1,1,0),(1,1,1),(1,0,1)$, giving $\tfrac29$, $\tfrac13$,
$\tfrac49$ and $\tfrac13$; they are unequal, so the origin of the chart,
$t_0=b/|b|^2$, is not the center of the quadrilateral.

At $\psi=0$ the ansatz is the exact solution, and we confirm this in both
forms of the equation. The root-mean-square departure of the residual from its
mean over $1024$ samples is $5.6\times10^{-15}$ in the cone form and
$7.3\times10^{-16}$ in the transverse form, with worst points at
$1.3\times10^{-13}$ and $2.4\times10^{-15}$, so the two agree
on this exact canonical-potential control. We also test an incorrect normalization. Unlike the cone form, the transverse form carries a
normalization the polytope does not fix, and setting $\Lambda=4$ in place of the
correct $6$, the transverse Einstein constant on the link, lifts the same
root-mean-square departure to $0.40$, roughly fifteen orders above the correctly normalized transverse value.

The network --- two hidden layers of width $16$, $337$ weights, on $1024$
samples --- is put through the same gate, a pass-or-fail entry test on $\Tone$
in the transverse form. From a deliberately perturbed start it must return to
the known solution, and the thresholds were fixed in advance: a held-out
residual below $10^{-6}$ and $\sup|\psi|$ below $5\times10^{-3}$. It passes.
The run we report ends after $1500$ Adam steps at
$\sup|\psi|=6.5\times10^{-5}$ with a held-out residual of $5.9\times10^{-7}$.

This residual is limited by the chosen training budget. Carrying the same run to
$12{,}000$ steps takes the held-out residual down to $3.55\times10^{-9}$,
still falling. The same test posed in the cone form ends its $1500$ steps at
$\sup|\psi|=1.57\times10^{-4}$ and $1.75\times10^{-6}$, which clears the
$\sup|\psi|$ threshold but misses the residual one. We therefore read the gate
as a pass-or-fail entry test and not as a measurement.

\subsection{\texorpdfstring{$Y^{2,1}$}{Y(2,1)}: a learned correction checked against GMSW}
\label{sec:ypq}

On $\Tone$ the answer was $\psi=0$ and the solver had only to stay there. Here
the correction is nonzero and unknown to it for the first time, and the closed
form is never shown to the optimizer. The cone over $Y^{2,1}$ is the complex
cone over $\dP1$, up to a $GL(2,\Z)$ change of basis, and its Reeb vector is
irregular: the orbits close only when $4p^2-3q^2$ is a perfect square
\cite[(3.8)]{Gauntlett:2004yd}, and here it is $13$. $\dP1$ carries no
K\"ahler--Einstein metric, and with $\dP2$ it is one of only two del Pezzo
surfaces that do not \cite{Tian:1990}. This target is therefore the situation of
section~\ref{sec:dp2} --- no K\"ahler--Einstein metric on the surface, a
Sasaki--Einstein metric on the link of the cone at an irregular Reeb vector --- with a
closed form to check against.

\begin{figure}[t]
\centering
\includegraphics[width=\textwidth]{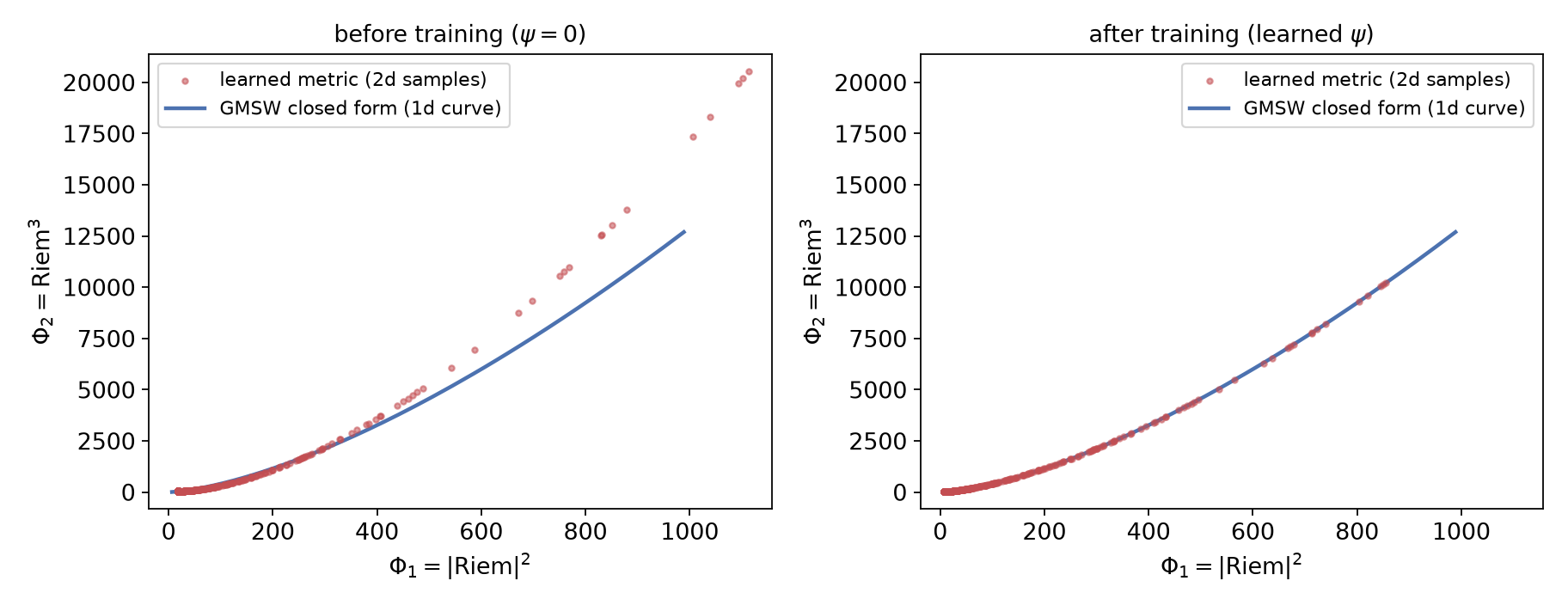}
\caption{$Y^{2,1}$ graded against the GMSW closed form in the plane of the two
curvature invariants, before training at $\psi=0$ (left) and after (right), at
the same $400$ sample locations. The solid curve is the closed form. Neither
axis requires a matching of coordinates, since both are scalar invariants. The
image is already close to one-dimensional before training, and the closed
form, being cohomogeneity one, lies exactly on a curve the ansatz was never
told of; the fit reduces the large-deviation tail and brings the image closer to
the exact curve.}
\label{fig:gmsw}
\end{figure}

After a degree-$8$ fit in the plain monomial basis --- raw, in the vocabulary
of section~\ref{sec:whiten}, which also improves on it --- the sample points collapse onto the closed-form
curve in the plane of the two curvature invariants (figure~\ref{fig:gmsw}). Over $400$ points the
median deviation falls from $27\%$ at $\psi=0$ to $0.15\%$. This grading needs
no matching of coordinates, since both axes are scalar invariants.

The check maps the two-dimensional slice into the plane by
$s\mapsto(\Phi_1,\Phi_2)$, so its image could fill a region; after training it
is a curve, as the cohomogeneity one of $\Ypq$ requires, every scalar
invariant being a function of the single coordinate the isometries leave,
and the ansatz, a free function on the polygon, was never told of it. Two
invariants falling on a curve is consistent with that symmetry and is not
offered as a proof of it. The learned correction is small against the
canonical potential, $\sup|\Gcan|=0.73$ on the same slice against
$\sup|\psi|=2.97\times10^{-2}$.

For a matched comparison on $Y^{2,1}$, we train both parametrizations with
Adam on the same samples. The polynomial reaches $5.14\times10^{-9}$ and the
network $5.17\times10^{-8}$; their gauge-fixed potentials agree to
$\sup_s|\psi_{\rm NN}-\psi_{\rm poly}|=2.6\times10^{-5}$. We use the polynomial
as a pointwise reference for the network where such comparisons are reported.

A separate network run uses an Adam warm-up followed by L-BFGS. It meets the
optimizer's convergence test after $5088$ iterations, with a held-out residual
of $1.333\times10^{-12}$. Its median and maximum deviations from the GMSW
curve are $0.0038\%$ and $0.0172\%$. The potential agrees with the polynomial
to $1.70\times10^{-7}$, compared with $\sup|\psi|=2.965\times10^{-2}$.

\subsection{Conditioning of the polynomial basis}
\label{sec:whiten}

The polynomial design matrix consists of basis functions evaluated at the
sample points. Its condition number $\kappa$, the ratio of its largest to its
smallest singular value, can be
computed before fitting. A large value makes some coefficient directions
difficult to resolve numerically and can cause a residual plateau.
Orthonormalizing the design matrix improves conditioning without changing
the function space.

We use Householder QR, whose computed orthogonality is accurate to machine
precision, independently of the condition number of the input. In contrast,
the loss of orthogonality in modified and classical Gram--Schmidt can grow as
$\kappa$ and $\kappa^2$, respectively
\cite{Higham:2002,TrefethenBau:1997}. The columns of $Q$ define the new basis.
Symmetry averaging introduces exact linear dependencies; in that case we use
an SVD with the rank fixed by the invariant count of
section~\ref{sec:dp3sym}. Raw-monomial and orthonormalized fits are
distinguished throughout. Further instances of rank loss are recorded in
appendix~\ref{app:robust}.

$Y^{3,2}$ is where we diagnose an ill-conditioned plateau, since the closed
form fixes what the answer should be. With raw
monomials the residual stalls at $6.0\times10^{-8}$ at degree $10$, and adding
degree $12$ makes it no better. The corresponding condition numbers are
$\kappa=4.3\times10^{8}$ at degree $10$ and $4.1\times10^{10}$ at degree $12$,
so the directions that higher degree adds are poorly resolved in the raw
basis.

Refitting the same function space in an orthonormal basis reaches
$2.0\times10^{-14}$. This reduction within the same function space identifies poor conditioning
as the cause of the earlier plateau. Orthonormalization also improves the
$Y^{2,1}$ fit. Keeping degree $8$ and
orthonormalizing alone takes the in-sample residual from $1.6\times10^{-9}$
down to $5.6\times10^{-13}$,
and to $6.1\times10^{-13}$ held-out, converging in $453$ iterations. The two
fits agree pointwise to $\sup_s|\psi_{\rm raw}-\psi_{\rm ortho}|=8.4\times10^{-6}$
against $\sup|\psi|=2.965\times10^{-2}$, and their fitted constants agree to six
figures, so the fitted functions agree closely. We compare them
pointwise rather than through $\sup|\psi|$ because that supremum carries the
affine gauge.

Independent geometric checks confirm the improvement. The GMSW deviation, the
distance from the closed-form curve of section~\ref{sec:ypq} and quoted
throughout as the median over the sample points followed by the maximum, falls
from $1.3\%/7.8\%$ to $0.008\%/0.04\%$ on
$Y^{3,2}$ and from $0.15\%/0.83\%$ to $0.007\%/0.02\%$ on $Y^{2,1}$, and in the
orthonormalized fits the fourth-order Abreu scalar deviates from $12$ by $0.00048\%$ on
average on $Y^{3,2}$ and by $0.0014\%$ on $Y^{2,1}$, with worst values of
$0.0078\%$ and $0.015\%$ on the $240\times240$ grid of
section~\ref{sec:graders}.

\subsection{A sweep in \texorpdfstring{$(p,q)$}{(p,q)}}
\label{sec:sweep}

We extend the test to five members of the family to assess variation
across $(p,q)$. Across the examples of table~\ref{tab:sweep} the held-out
residual runs from $8.6\times10^{-17}$ to $1.7\times10^{-12}$; all five in the
orthonormalized basis of section~\ref{sec:whiten}. The
column $\delta_y=3q/2p$ is the separation $y_2-y_1$ of two roots of the closed
form \cite[(3.1)]{Gauntlett:2004yd}, recorded in appendix~\ref{app:ypq}, and it
determines the metric. The one quasi-regular member, $(7,3)$, gives the
smallest residual, more than two orders of magnitude below the best of the two
irregular members fitted at the same degree.

\begin{table}[htbp]
\centering
\footnotesize
\begin{tabular}{lcccccc}
\toprule
$(p,q)$ & $\delta_y$ & class & deg & held-out & GMSW med/max & Abreu mean/worst\\
\midrule
$(2,1)$ & $0.750$ & irregular & 8 & $6.1\times10^{-13}$ & $0.007\%/0.02\%$ &
$0.0014\%/0.015\%$\\
$(3,2)$ & $1.000$ & irregular & 12 & $2.0\times10^{-14}$ & $0.008\%/0.04\%$ &
$0.00048\%/0.0078\%$\\
$(4,3)$ & $1.125$ & irregular & 12 & $1.7\times10^{-12}$ & $0.045\%/0.27\%$ &
$0.0043\%/0.047\%$\\
$(5,4)$ & $1.200$ & irregular & 14 & $6.6\times10^{-13}$ & $0.042\%/0.22\%$ &
$0.0031\%/0.054\%$\\
$(7,3)$ & $0.643$ & quasi-regular & 12 & $8.6\times10^{-17}$ &
$0.0025\%/0.0075\%$ & $6.0\times10^{-6}\,\%/2.9\times10^{-4}\,\%$\\
\bottomrule
\end{tabular}
\caption{The sweep across $Y^{p,q}$, all five members in their orthonormalized
fits. The Abreu column is the estimator of section~\ref{sec:graders}: the mean
deviation from $S=12$, and beside it the worst value on the $240\times240$ grid
it was taken on.}
\label{tab:sweep}
\end{table}

\section{The \texorpdfstring{$\dP3$}{dP3} hexagon: a published numerical reference}
\label{sec:dp3}

\subsection{\texorpdfstring{$\dP3$}{dP3} as the positive control}
\label{sec:dp3why}

For the third target a published computation is available. $\dP3$ carries a
K\"ahler--Einstein metric \cite{TianYau:1987,Siu:1988,Tian:1990}. DHHKW solved that same
problem three ways --- Ricci flow in complex coordinates, Ricci flow in
symplectic coordinates, and a constrained optimization --- and report the three
metrics agreeing to one part in $10^6$ \cite{Doran:2007zn}. The quantities we
compare against below are from the third, and
their smooth part $h_{\rm D}$ is our $\psi$ up to the chart and the
normalization, $h_{\rm D}(\tilde s)=3\psi(\tilde s/3)$ modulo an affine function. They expand
it, as we do, in $D_6$-invariant polynomials.

$\dP3$ is therefore both the target with a published computation and the positive control.
It is the only target with an external reference and no closed form, which is
exactly the situation the whole validation is built for. At a regular Reeb
vector the leaf space is a smooth surface, so one unknown function carries
three roles at once. The slice potential is the potential of the
K\"ahler--Einstein metric on that surface, and the same function is the
transverse part of the five-dimensional Sasaki--Einstein metric and of the
six-dimensional Ricci-flat cone.

\subsection{A \texorpdfstring{$D_6$}{D6}-invariant polynomial ansatz}
\label{sec:dp3sym}

Throughout, the degree of a polynomial means its total degree in the slice
coordinates $s$; the display chart $\tilde s=3s$ rescales them and does not change it.
Following DHHKW \cite[(5.1) and Appendix~C]{Doran:2007zn}, we use the generators
\begin{equation}
U=\tilde s_1^2+\tilde s_1\tilde s_2+\tilde s_2^2,
\qquad V=\tilde s_1^2\tilde s_2^2(\tilde s_1+\tilde s_2)^2.
\label{eq:UV}
\end{equation}
The monomials $U^iV^j$ with $2\le2i+6j\le D$ give $14$ coefficients at
$D=14$ and $21$ at $D=18$, the latter being DHHKW's $22$ coefficients
with the constant removed \cite[\S6]{Doran:2007zn}.

How much this saves is set by the order of the symmetry group, which is why it
is worth doing on the hexagon and not on the pentagon. At degree $14$ there are $117$ polynomials once
the three gauge directions are dropped. The hexagon's $D_6$ has twelve elements
and leaves $14$ of them, a factor $8$; the fan of the lattice pentagon has exactly two lattice
automorphisms, the identity and one reflection --- a five-fold rotation is
already excluded by the crystallographic restriction --- and that reflection
leaves $62$, a factor $1.9$. The symmetry is
therefore imposed on the $\dP3$ polynomial, withheld from the $\dP3$ network so
that we can ask whether it is found rather than assumed, and not used on
$\dP2$. Imposing a symmetry at all
is justified by
uniqueness \cite{BandoMabuchi:1987}, since the solution really does have the
symmetry.

\begin{figure}[t]
\centering
\includegraphics[width=0.9\textwidth]{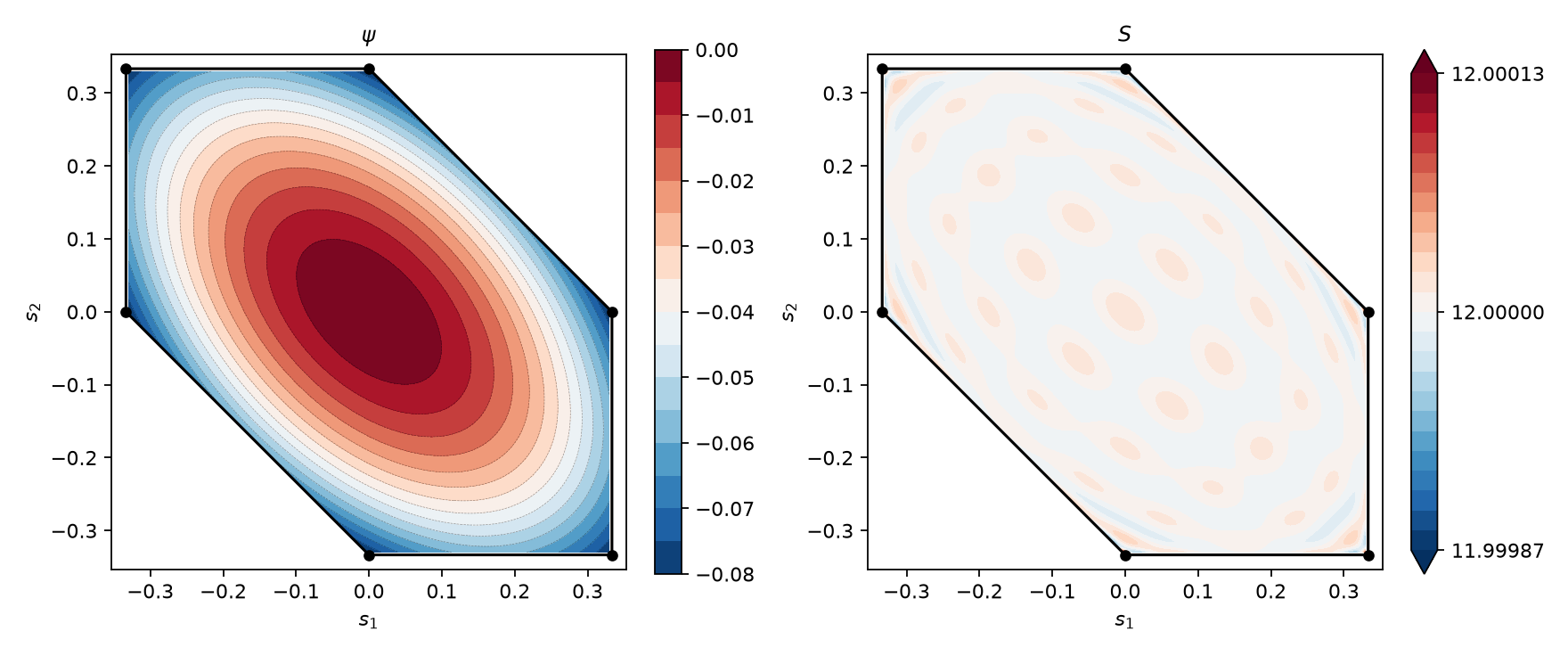}
\caption{The learned $\dP3$ correction $\psi$ on the hexagonal slice, in the
coordinates $s_1,s_2$. The function is manifestly invariant under the hexagon's
$D_6$, which is imposed here, and its maximum
$\sup|\psi|=8.13\times10^{-2}$ is attained at the six vertices. Right: the
Abreu scalar of the same fit on the same slice, on a colour scale spanning
$12\pm1.3\times10^{-4}$; values outside this interval use the endpoint
colours. The reference value $S=12$ is fixed in section~\ref{sec:graders}. That check is fourth order in $\psi$ where the objective is second,
and it enters no objective.}
\label{fig:dp3learned}
\end{figure}

Symmetrization also removes the gauge without a separate construction for
two reasons. The two linear flat directions cannot arise
because no linear function is $D_6$-invariant; the constant cannot because the
invariant basis starts at degree two. The fit itself is
small and fast. Fourteen invariant coefficients at degree $14$ reach a
held-out residual of $9.1\times10^{-14}$ in $161$ iterations, with
$\sup|\psi|=8.13\times10^{-2}$ attained at the six vertices
(figure~\ref{fig:dp3learned}). The rank of the averaged basis is fixed against
these invariant counts rather than by a tolerance
(appendix~\ref{app:robust}). Escalating to degree $18$ gives $21$
invariants and a held-out residual of $1.0\times10^{-16}$.

\subsection{Comparison with DHHKW's metric}
\label{sec:dp3grading}

Comparison with DHHKW needs a normalization bridge,
\begin{equation}
S_{\rm DHHKW}=\tfrac13\,S=4,
\qquad
\lambda_{\rm DHHKW}=\tfrac13\,\lambda .
\label{eq:dhhkwnorm}
\end{equation}
DHHKW's flow fixes $\Ric=g$ on the transverse surface, whereas our doubled slice has
$\Ric(\tilde g_T)=3\tilde g_T$, and \eqref{eq:dhhkwnorm} is that factor of three carried into the
scalar curvature and the eigenvalues; on the potential it reads
$G_{\rm DHHKW}(\tilde s)=3\,G_P(\tilde s/3)$ up to an affine function, with $\tilde s=3s$ the chart
of figure~\ref{fig:polytopes}, hence $h_{\rm D}(\tilde s)=3\psi(\tilde s/3)$. Their eigenvalues are
those of eigenfunctions invariant under $D_6$ \cite[\S6.1]{Doran:2007zn}, and
ours are computed in the same sector, in the invariant basis of
section~\ref{sec:dp3sym}; they are not the lowest torus-invariant
eigenvalues. Those belong to the symplectic coordinates themselves, which their
eq.~(3.40) makes eigenfunctions of $-\triangle$ with eigenvalue $2\Lambda$, that
is $2$ in their units and $12$ on the link, and table~\ref{tab:fullsector} lists
the low torus-invariant spectrum with its $D_6$ labels: $6.3228$ is the third
level above the constant, and the lowest of those that are $D_6$-invariant. In their normalization and sector we find
$\lambda_1^{D_6}=6.3228$ against their
$6.322$, $\lambda_2^{D_6}=17.094$ against their $17.2$, and a leading eigenfunction
coefficient of $-0.244$ against their $-0.245$, that coefficient being their
$X_1$, the weight of the invariant $U$ in their expansion of the
eigenfunction, compared in their normalization $\psi_1(0)=0.1$ of
\cite[(6.7)]{Doran:2007zn} with $U$ of \eqref{eq:UV} in the chart $\tilde s$. DHHKW's tabulated results indicate that
$\lambda_2$ has not converged: it still moves as $17.4\to17.2$ across the two orders they
compute and its fit residual is $270$ times worse than that of their converged
$\lambda_1$. Our value near $17.094$ is an estimate from the finite
approximation specified in table~\ref{tab:fullsector}; we do not assign a
convergence error to its last quoted digits.

\begin{table}[htbp]
\centering
\footnotesize
\begin{tabular}{rrcl}
\toprule
level & $\lambda$ ($S=4$) & degeneracy & $D_6$ irrep\\
\midrule
1 & $2.00000$ & 2 & $\mathrm{E}_1$ (the symplectic coordinates)\\
2 & $4.74276$ & 2 & $\mathrm{E}_2$\\
3 & $6.32277$ & 1 & $\mathrm{A}_1$ (DHHKW's $\lambda_1$)\\
4 & $7.23816$ & 1 & $\mathrm{B}_1$\\
5 & $9.58011$ & 1 & $\mathrm{B}_2$\\
6 & $11.15026$ & 2 & $\mathrm{E}_1$\\
7 & $12.31061$ & 2 & $\mathrm{E}_2$\\
8 & $17.09405$ & 1 & $\mathrm{A}_1$ (DHHKW's $\lambda_2$)\\
9 & $17.70172$ & 2 & $\mathrm{E}_1$\\
10 & $17.98978$ & 2 & $\mathrm{E}_2$\\
\bottomrule
\end{tabular}
\caption{The low torus-invariant spectrum of $-\triangle$, zero mode excluded, on
the degree-$18$ $\dP3$ metric, in a general polynomial basis of degree $16$ with
the deterministic quadrature. The entries are computed values, not certified
accuracies. Eigenvalues are in DHHKW's normalization $S=4$; by
\eqref{eq:dhhkwnorm} the value on our slice is three times the entry, and by
section~\ref{sec:graders} the value on the link at $r=1$ is six times it. The
representation is read off the character of each eigenspace, every character
coming out an integer to $10^{-3}$. The labels are the standard ones for $D_6$:
$\mathrm{A}$ and $\mathrm{B}$ are one-dimensional and $\mathrm{E}$ two-dimensional; on an order-six
generator the character is $+1$ in $\mathrm{A}$ and $\mathrm{E}_1$ and $-1$ in $\mathrm{B}$ and $\mathrm{E}_2$; and the
subscript on $\mathrm{A}$ and $\mathrm{B}$ records the sign under the three reflections that fix a
vertex, $+1$ for $\mathrm{A}_1$ and $\mathrm{B}_1$. The first level is the pair of symplectic
coordinates at exactly $2\Lambda$; the values compared with DHHKW are the two
lowest levels of the trivial representation $\mathrm{A}_1$.}
\label{tab:fullsector}
\end{table}

\begin{figure}[t]
\centering
\includegraphics[width=0.8\textwidth]{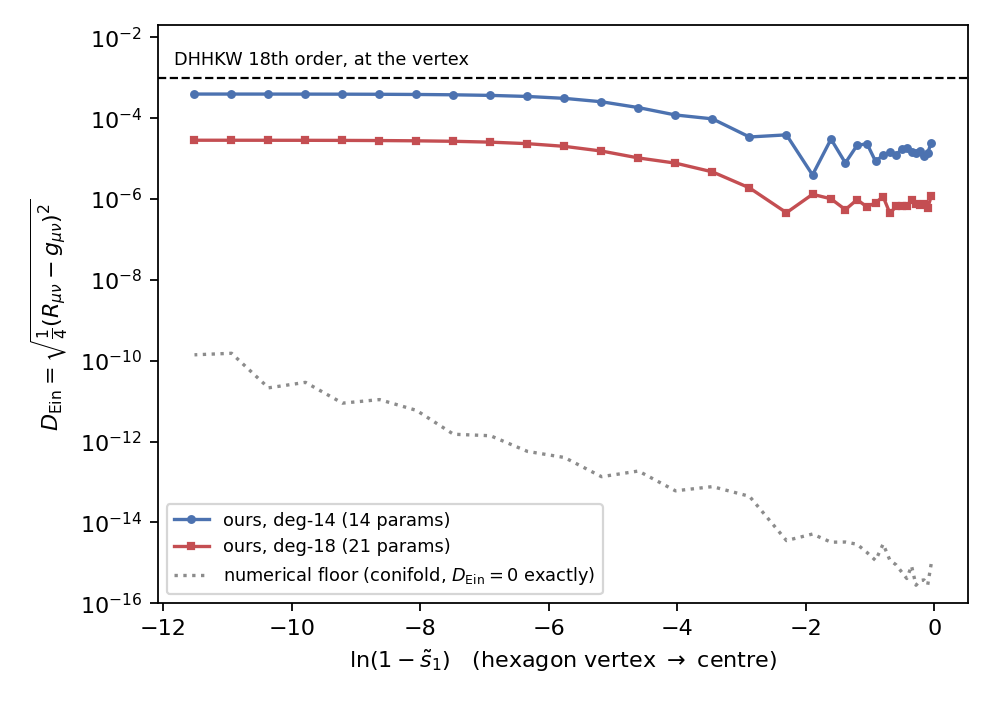}
\caption{DHHKW's pointwise scale $D_{\rm Ein}$ along the ray from the center to a vertex
of the hexagon, for the degree-$14$ and degree-$18$ fits. The dashed line is
DHHKW's own $\approx10^{-3}$ at the vertex. The dotted curve is the floor of
the numerical evaluation, measured on the conifold where $D_{\rm Ein}$ vanishes identically;
it is not a constant, running from $1.5\times10^{-10}$ at the vertex down to
$10^{-16}$ near the centre.}
\label{fig:dprofile}
\end{figure}

Throughout this subsection $D_{\rm Ein}$ is DHHKW's pointwise measure of how far a metric
is from the Einstein condition,
\begin{equation}
D_{\rm Ein}=\sqrt{\tfrac14\bigl(R_{\mu\nu}-g_{\mu\nu}\bigr)\bigl(R^{\mu\nu}-g^{\mu\nu}\bigr)}\,,
\label{eq:Dmeasure}
\end{equation}
their eq.~(5.4), in the normalization $\Ric=g$ that their flow fixes. It scores
a metric against the equation and not against another metric, so no reference
metric enters and it vanishes identically on an exact solution --- which is
what makes the conifold value quoted below a noise floor rather than a discrepancy.
Our convention has $\Ric=3g$; since $\Ric(cg)=\Ric(g)$, the rescaled $g'=3g$
satisfies $\Ric(g')=g'$ and \eqref{eq:Dmeasure} applies to it verbatim. We
evaluate $D_{\rm Ein}$ along the vertex ray, the segment from the centre of the hexagon
to one of its vertices, which is where two facets meet, the canonical part of
the potential is most singular, and the higher derivatives of a polynomial
correction are least constrained by the samples.
Along that ray $D_{\rm Ein}$ is largest at the vertex, where it reaches
$3.94\times10^{-4}$ at degree $14$ and $2.84\times10^{-5}$ at degree $18$,
against the $\approx10^{-3}$ that DHHKW's eighteenth-order expansion reaches
at the same point \cite[Fig.~12]{Doran:2007zn}.
That expansion comes from DHHKW's constrained-optimization method, the one of
their three that shares our variable and our parametrization, so this is a
comparison within one representation and not across two. On the exact conifold, the same evaluation gives at most
$1.5\times10^{-10}$ at the vertex and falls to $10^{-16}$ near the centre.
This numerical baseline lies five to six orders of magnitude below the
$\dP3$ values in figure~\ref{fig:dprofile}.

DHHKW's harmonic $(1,1)$-forms provide a further check of the metric
\cite{Doran:2007zn}, with the conversions of appendix~\ref{app:conventions}.
We quote their printed potential as
\[
\mu_a^{\rm src}=\ln(1+w_a\cdot \tilde s)+\sum c_{nm}\tilde s_1^n\tilde s_2^m .
\]
In our real-form convention, \(\mu_a=-2\mu_a^{\rm src}\) and
\(\theta_a=i\partial\bar\partial\mu_a\), with the conversion calibrated in
table~\ref{tab:conventions} of appendix~\ref{app:conventions}, where the
chain is derived. The logarithm in their expression
\cite[(6.11)]{Doran:2007zn} fixes the polytope to \(1+w_a\cdot \tilde s\ge0\),
so their chart is \(\tilde s=3s\).

Harmonicity requires \(\triangle\mu_a\) to be constant
\cite[(3.46)]{Doran:2007zn}. The contraction in DHHKW's metric normalization is
\begin{equation}
g_{\rm DHHKW}^{i\bar\jmath}(\theta_a)_{i\bar\jmath}
=-\tfrac13\triangle\mu_a^{\rm src}
=\tfrac16\triangle\mu_a ,
\label{eq:contraction}
\end{equation}
where \(\triangle\) is the transverse Laplacian of
section~\ref{sec:graders}. The exact contractions are
\(\tfrac23\) for \(\theta_2\) and \(2\) for
\(\omega=\tfrac12\sum_a\theta_a\)
\cite[(6.14)--(6.15)]{Doran:2007zn}: the dihedral symmetry of $\dP3$ makes the contraction of \(\theta_a\) independent of \(a\).
DHHKW's fitted value \(0.6672\) is reproduced as \(0.667037\) on the fitting domain.

The domain mean of the contraction \eqref{eq:contraction} is in fact
independent of the smooth metric
correction. By the divergence theorem, the integral of
\(\triangle\mu_a^{\rm src}\) is a boundary term. The vanishing inverse Hessian
cancels the logarithmic derivative at the facet, leaving a value fixed by the
Guillemin asymptotics; the smooth polynomial part contributes no boundary
term. Deterministic integration gives \(\tfrac23\) for the mean contraction
of \(\theta_2\), and \(2\) for that of \(\omega\), on both the fitted metric and
the untrained Guillemin metric to \(10^{-13}\). The agreement therefore verifies
our convention and the normalization rather than the numerical solution itself.

The spatial variation does depend on the metric. The constancy residuals
quoted here are rms spreads of the contraction \eqref{eq:contraction} itself,
the quantity whose exact
value is \(\tfrac23\) and whose mean is quoted above, which is also the quantity
for which DHHKW report an rms of about \(10^{-3}\) on their fitting domain
\cite[(6.12)]{Doran:2007zn}.
The spread is \(6.129\times10^{-3}\) on our metric and
\(3.568\times10^{-1}\) on Guillemin, a ratio of \(58.2\). We call this ratio
the discrimination.

The Guillemin comparison keeps the published
scalar potential fixed and reconstructs its two-form with each metric.
Tables~\ref{tab:oneone} and~\ref{tab:dhhkw} summarize these checks.

\begin{table}[htbp]
\centering
\footnotesize
\begin{tabular}{lccc}
\toprule
measure on the $(1,1)$-form & our metric & Guillemin & discrimination\\
\midrule
published constant, $|{\cdot}-\tfrac23|$ & $<10^{-13}$ & $<10^{-13}$ &
protected\\
constancy, rms spread & $6.129\times10^{-3}$ & $3.568\times10^{-1}$ &
$\times58.2$\\
$\omega$ anchor, $|{\cdot}-2|$ & $<10^{-13}$ & $<10^{-13}$ &
protected\\
\bottomrule
\end{tabular}
\caption{Measures on DHHKW's published $(1,1)$-form, evaluated on our metric and on the untrained Guillemin potential over the whole hexagon, integrated by the deterministic rule of section~\ref{sec:graders}. The discrimination is the ratio of the two entries in the row. Both means are boundary integrals and therefore exactly protected, and the $\omega$ anchor is not independent of the first row, following from it by $\omega=\tfrac12\sum_a\theta_a$ and the dihedral symmetry; only the spread is a measurement, and a sample estimate of either mean returns its own sampling error in place of a discrimination.}
\label{tab:oneone}
\end{table}

\begin{table}[htbp]
\centering
\footnotesize
\begin{tabular}{lll}
\toprule
quantity & DHHKW & this work (deg 14 / 18)\\
\midrule
$\lambda_1^{D_6}$ ($S=4$) & $6.322$ & $6.3228$\\
$\lambda_2^{D_6}$ ($S=4$) & $17.2$ & $17.094$\\
eigenfunction coefficient & $-0.245$ & $-0.244$\\
$D_{\rm Ein}$ on the vertex ray & $\approx10^{-3}$ & $3.94\times10^{-4}$ /
$2.84\times10^{-5}$\\
\bottomrule
\end{tabular}
\caption{The comparison with DHHKW, quantity by quantity. Eigenvalues are the two lowest of the $D_6$-invariant sector, in their normalization $S=4$. The constancy residuals are not here because neither of the two is a published number of theirs; both are measured on our metric, and table~\ref{tab:oneone} carries them.}
\label{tab:dhhkw}
\end{table}

\subsection{Recovery of unimposed \texorpdfstring{$D_6$}{D6} symmetry}
\label{sec:nnpos}

The invariant polynomial ansatz imposes $D_6$ symmetry. To test its recovery
without an architectural constraint, we fit a symmetry-free network, one
without symmetry averaging. We measure the \emph{equivariance defect}

\begin{equation}
\max_{g\in D_6}\frac{\lVert H_\psi(gs)-g^{-\mathrm{T}}H_\psi(s)g^{-1}\rVert}{\lVert H_\psi(s)\rVert},
\qquad H_\psi=\Hess\psi ,
\label{eq:equivar}
\end{equation}
averaged over the held-out samples, with $\lVert\cdot\rVert$ the Frobenius norm
in the chart $s$ --- the $D_6$ matrices are not orthogonal there, so the size of
the defect depends on the chart while its vanishing does not. This is the equivariance error of the
equivariant-network literature \cite{Weiler:2019e2cnn,Gruver:2023lie}, the
relative norm of $f(g\cdot x)-\rho(g)f(x)$, specialized to $f=\Hess\psi$ with
$\rho$ the transformation of a covariant two-tensor and the maximum taken over
the finite group. The Hessian annihilates affine functions, so the
measure does not see the gauge.

\begin{figure}[t]
\centering
\includegraphics[width=0.7\textwidth]{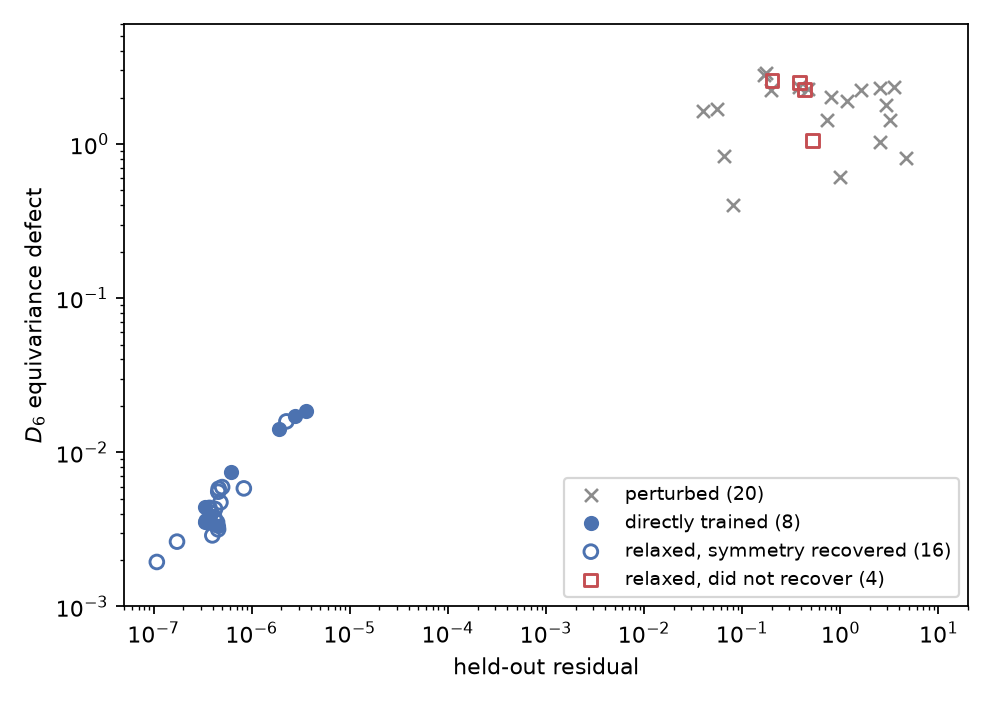}
\caption{Every symmetry-free network state of the $\dP3$ study, in the plane
of held-out residual against equivariance defect. Filled circles are the eight
directly trained runs. Crosses are perturbed states, four seeds at each of
$\epsilon=0.10$, $0.25$, $0.50$, $0.75$ and $1.00$, which carry a defect of
order one. Open circles and squares are those same states after relaxation on
the Monge--Amp\`ere residual alone, with no symmetry term anywhere in the
objective: the circles have returned to the converged cluster and the squares
have not. These runs demonstrate symmetry recovery without an explicit symmetry
penalty.}
\label{fig:defectlaw}
\end{figure}

The measure is tested in both directions. Over an eight-seed ensemble of
symmetry-free networks the initial defect averages $2.509$, since generic weights carry no symmetry, and
falls after training to $9.07\times10^{-3}\pm6.5\times10^{-3}$, a collapse of
$\times277$. Across the same seeds the held-out residual itself spreads by
$\times10.7$, a dependence the polynomial does not have. We also test whether symmetry is recovered after perturbing a converged
network. We take a converged state, perturb every weight by a relative
amount $\epsilon$, which destroys the symmetry and returns the defect to order
one, and then resume training on the Monge--Amp\`ere residual alone, with no
symmetry term anywhere in the objective. We call the second step \emph{relaxation}.
Four seeds were relaxed at each of five sizes, twenty runs in all. We count a
run as recovered when the defect falls back from order one to the level of the
directly trained states, and as failed when it stays of order one. Recovery
falls off with the size but does not stop: all four seeds recover at
$\epsilon=0.10$ and at $0.25$, three of four at $0.50$ and at $0.75$, and two of
four at $\epsilon=1.00$. Recovery becomes less frequent with increasing perturbation,
but no sharp threshold is resolved by these twenty runs.
Figure~\ref{fig:defectlaw} shows every state of this study in the plane of
held-out residual against defect: the eight directly trained runs, the twenty
perturbed states at defect of order one, and the same twenty after relaxation.
Fourteen of the sixteen recovered states land below the directly trained ones
in both residual and defect, and the four that fail keep residual and defect
large together.

We next compare three parametrizations with different symmetry constraints. The polynomial, the $D_6$-informed network --- a multilayer perceptron
evaluated at the twelve images of the input point under $D_6$ and its outputs
averaged, so that the symmetry is exact by construction --- and the
symmetry-free network approximate the metric (table~\ref{tab:threeway}), at held-out residuals of
$9.1\times10^{-14}$, $8.8\times10^{-8}$ and $4.8\times10^{-7}$. The two network
numbers come from production runs at different iteration caps, neither at
its floor; carried to their own convergence tests, the $D_6$-informed
network reaches $3.3\times10^{-10}$ and the symmetry-free one
$1.16\times10^{-8}$, so withholding the symmetry raises the held-out
residual by a factor $35$ and the budget by a factor $75$
(appendix~\ref{app:robust}).

\begin{table}[htbp]
\centering
\footnotesize
\begin{tabular}{lccc}
\toprule
$\dP3$ & polynomial & network, $D_6$ & network, no $D_6$\\
\midrule
held-out residual & $9.1\times10^{-14}$ & $8.8\times10^{-8}$ &
$4.8\times10^{-7}$\\
Abreu $S=12$, worst/mean & $0.019\%/0.001\%$ & $1.4\%/0.22\%$ &
$8.1\%/0.54\%$\\
$\lambda_1^{D_6}$ ($S=4$) & $6.3228$ & $6.3220$ & $6.3232$\\
\bottomrule
\end{tabular}
\caption{The same positive control run three ways, at the production caps of table~\ref{tab:cost}. The Abreu scalar is in our normalization $S=12$, the eigenvalue in DHHKW's $S=4$.}
\label{tab:threeway}
\end{table}

\section{The \texorpdfstring{$\dP2$}{dP2} cone: construction and negative controls}
\label{sec:dp2}

\subsection{Two problems on one polygon}
\label{sec:dp2why}

The $\dP2$ pentagon carries two targets that differ only in the Reeb
vector. At the regular one the sole reference is a theorem, which forbids a
solution; at the volume-minimizing one a solution exists and there is no
reference at all. The theorem is Matsushima's \cite{Matsushima:1957}: a
K\"ahler--Einstein manifold has a reductive automorphism group, and the
automorphism group of $\dP2$ is not reductive. Among the del Pezzo surfaces
only $\dP1$ and $\dP2$ are excluded in this way \cite{Tian:1990}, and for toric
Fano surfaces the criterion is exact, a K\"ahler--Einstein metric existing if
and only if the Futaki invariant vanishes \cite{WangZhu:2004,Mabuchi:1987}.
$\dP2$ at the regular Reeb vector is therefore the negative control.

That obstruction has an equivalent formulation. The Futaki invariant is the first variation of the MSY
volume functional \eqref{eq:volmin} with respect to the Reeb vector
\cite{Martelli:2006yb}, and that functional is strictly convex with a unique
critical point, so the invariant vanishes at $b$ precisely when $b$ is the
minimizer. ``No K\"ahler--Einstein metric at the regular Reeb vector'' and
``the minimizer is not the regular Reeb vector'' are therefore the same
statement, and the negative control below is an instance of solving at a
displaced Reeb vector rather than a separate phenomenon. The pentagon offers almost
no symmetry to exploit: its fan has exactly two lattice automorphisms, the
identity and one reflection, $\Z_2$, a five-fold rotation being excluded by the
crystallographic restriction. We do not impose it,
since it would reduce the parameter count by only a factor $1.9$.

\subsection{The Sasaki--Einstein metric at \texorpdfstring{$b^{\ast}$}{b*}: a new construction}
\label{sec:dp2se}

At its other Reeb vector the same pentagon carries a solvable problem. At the
irregular minimizer $b^{\ast}$ the link of the cone admits a Sasaki--Einstein
metric \cite{Futaki:2006cc}, and the two-point blow-up of $\CP^2$ is the worked
example of that paper. No closed form is known for it, the cone is not a member
of $L^{a,b,c}$, and DHHKW deferred it, so unlike every earlier target there is
no independent metric to compare with; the one external datum is an integrated
curvature invariant, used at the end of this subsection. The irrationality of $b^{\ast}$
changes nothing in the equation and calls for no modification of the method.
What it removes is the transverse surface: there is no quotient, so the slice
potential is no longer the potential of a K\"ahler--Einstein surface, and of
the three roles that one function carried on the hexagon in
section~\ref{sec:dp3why} only the five-dimensional link and the
six-dimensional cone survive.

\begin{figure}[t]
\centering
\includegraphics[width=\textwidth]{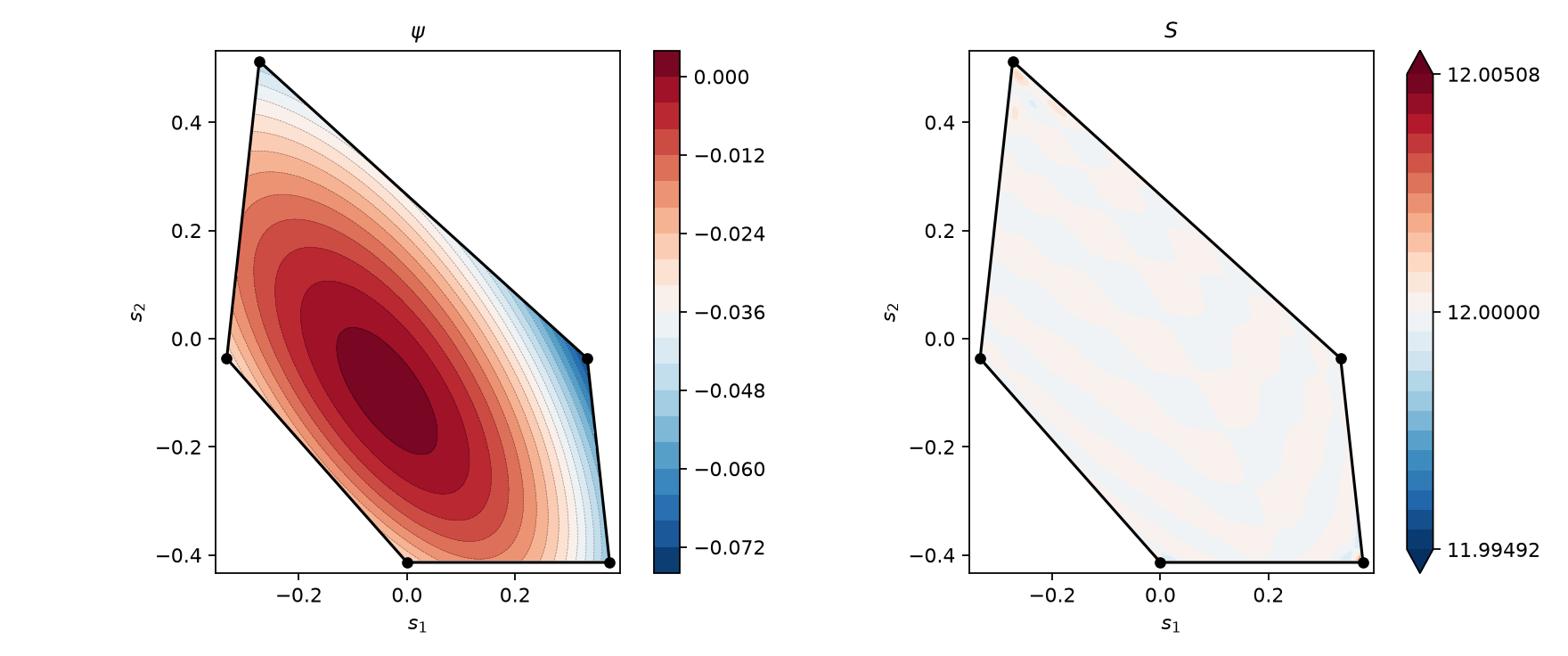}
\caption{The $\dP2$ correction $\psi$ at the irregular minimizer $b^{\ast}$ on
the pentagonal slice, and the Abreu scalar of the same degree-$14$ fit. Left:
$\psi$ runs from $-7.29\times10^{-2}$ to $2.21\times10^{-3}$ and attains its
extreme value at one vertex, the pentagon having only the $\Z_2$ of
section~\ref{sec:dp2why}, which is not imposed. Right: $S$ on a colour scale of
width $5.1\times10^{-3}$ about the value $S=12$ that
section~\ref{sec:graders} requires, the deviation over this grid being
$0.0368\%$ at its worst and $0.0006\%$ in the mean. That check is fourth order
in $\psi$ where the objective is second, and it enters no objective. The
counterpart for the hexagon is figure~\ref{fig:dp3learned}.}
\label{fig:dp2learned}
\end{figure}

The construction then runs exactly as on every solvable target above. The
held-out residual falls monotonically from $1.20\times10^{-7}$ at degree $6$ to
$6.04\times10^{-14}$ at degree $14$ and $1.5$--$3.4\times10^{-15}$ at degree
$16$, the last being the range over three independent sampling draws, with
L-BFGS meeting its convergence test at every degree; that sequence is the lower
branch of figure~\ref{fig:neg}.
The Abreu scalar \eqref{eq:abreu},
fourth order in the potential and never in the objective, comes out $S=12$ to
$0.0006\%$ in the mean and $0.0368\%$ at its worst on the $240\times240$ grid
of section~\ref{sec:graders}, and the network on the same target reaches $4.1\times10^{-10}$ when
its optimizer budget is released, short of the polynomial as everywhere else,
so both classes attain small residuals there. Figure~\ref{fig:dp2learned}
shows the correction and that check on the slice. To our knowledge, this metric has not previously been computed.

Computing a curvature invariant provides a nontrivial test of our numerical
result and of a conjecture that is unproved at irregular Reeb vectors. Eager, Gary and Roberts \cite{Eager:2010dk} proposed that for a
Sasaki--Einstein five-manifold $Y$ the integral $\int_Y|\mathrm{Riem}|^2$ is fixed by the
asymptotics of the Hilbert series of the cone. Arabi Ardehali, Liu
and Szepietowski \cite{ArabiArdehali:2014otj,ArabiArdehali:2015iow} gave the
toric form of the proposal, expressing the integral through the R-charges of the
extremal mesons, equivalently, on the geometry side, through the Reeb vector and
the generators of the dual cone. Nakayama and Okazaki \cite{Nakayama:2026bls} rederived it from the
residue of a supersymmetric zeta function and evaluated it for the del Pezzo
links; for $\dP2$, in the normalization $\Ric(g_Y)=4g_Y$ of
section~\ref{sec:setupgeom}, it gives
\begin{equation}
\int_Y|\mathrm{Riem}|^2=\tfrac{4}{243}\bigl(2551+295\sqrt{33}\bigr)\pi^3
=2166.941135\ldots
\label{eq:riem2dp2}
\end{equation}
For regular $Y$ the proposal was proved in \cite{Eager:2010dk} by reduction to
the K\"ahler--Einstein base; at an irregular Reeb vector it remains a conjecture.
We evaluate the integral on our metric from the cone invariant $\Phi_1$ of
\eqref{eq:inv6}: for a metric cone over an Einstein link with $\Ric(g_Y)=4g_Y$
the Riemann tensor of the cone is that of the link minus the constant-curvature
term, so that at $r=1$
\begin{equation}
|\mathrm{Riem}_Y|^2=\Phi_1\big|_{r=1}+40,
\label{eq:riem2cone}
\end{equation}
the constant being $2n(n-1)$ for $n=5$; the cone over the round $S^5$ is flat and
gives the known $40\pi^3$. On the degree-$14$ metric of this subsection the integral is
$2166.941140$, within $2.1\times10^{-9}$ of \eqref{eq:riem2dp2}; the same
evaluation reproduces the tabulated values for $\Tone$, where the integrand is
the constant $136$, and for $Y^{2,1}$.

\subsection{Residual plateaus at the regular Reeb vector}
\label{sec:fingerprint}

\begin{figure}[t]
\centering
\includegraphics[width=0.68\textwidth]{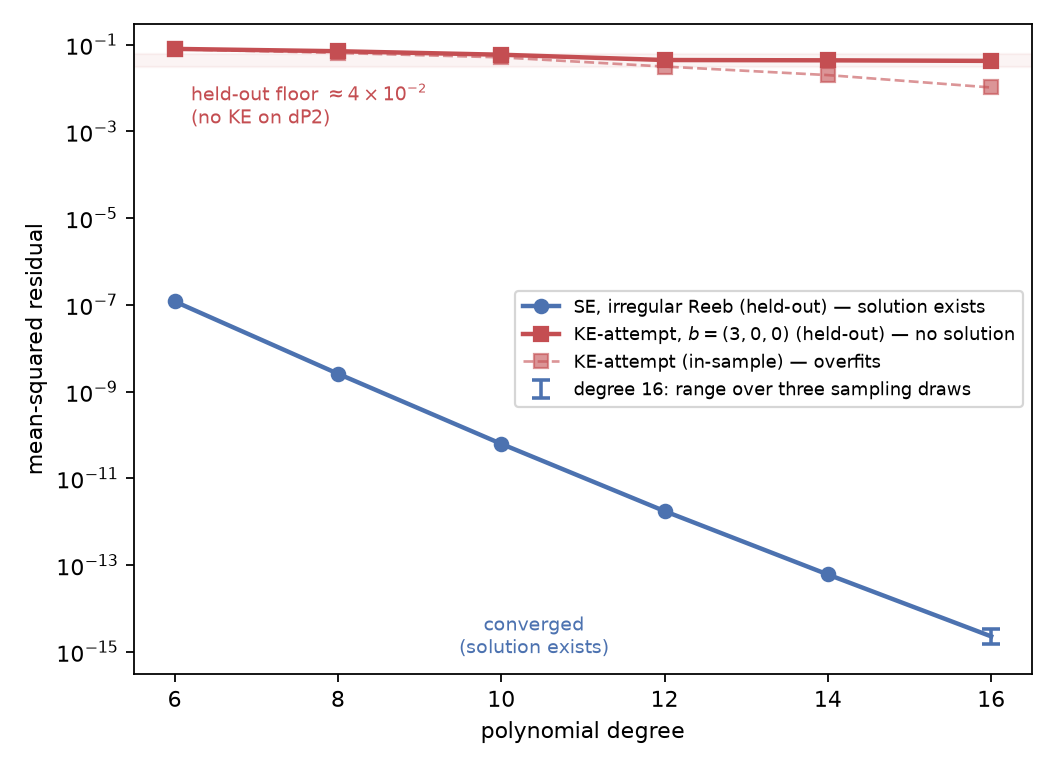}
\caption{One pentagon, two Reeb vectors, and the same machinery on both. At
the volume-minimizing irregular $b^{\ast}$ the held-out residual falls
monotonically to $6.0\times10^{-14}$ at degree $14$ and to
$1.5$--$3.4\times10^{-15}$ at degree $16$, where the bar is that range over
three independent sampling draws and the marker its geometric centre. At the
regular Reeb vector it floors at
$\approx4\times10^{-2}$ while the in-sample residual keeps falling, and that
separation is what the text calls the fingerprint: the behaviour observed where a
theorem forbids a solution.}
\label{fig:neg}
\end{figure}

We now run the full construction at the regular Reeb vector exactly as if a
solution existed. The held-out residual stalls at $\approx4\times10^{-2}$ and
stays there, while the same pentagon at the solvable Reeb vector reaches
$6\times10^{-14}$, twelve orders below. The optimizer corroborates the
reading, since L-BFGS never meets its convergence test here and exhausts its
evaluation budget --- \texttt{scipy}'s default of $15{,}000$ function and
gradient evaluations --- after some $13{,}700$ iterations (figure~\ref{fig:neg}).

On the obstructed target the in-sample residual decreases with degree while
the held-out residual remains large. We use \emph{fingerprint} only as a label
for this observed pattern. It is not specific to non-existence though: insufficient
sampling, loss of regularity at the boundary or optimization failure can also
produce overfitting. A criterion for non-existence would have to bound the infimum of
the residual from below, and the absence of a smooth solution does not imply
that this infimum is positive: a minimizing sequence may drive the residual
down while itself degenerating.

The comparison at $b^{\ast}$ shows that the implementation attains small
residuals on the same toric cone for a solvable Reeb vector. At the regular
vector, the polynomial plateau is robust under the tested changes of sample
draw, initialization and software build.

\begin{figure}[t]
\centering
\includegraphics[width=0.7\textwidth]{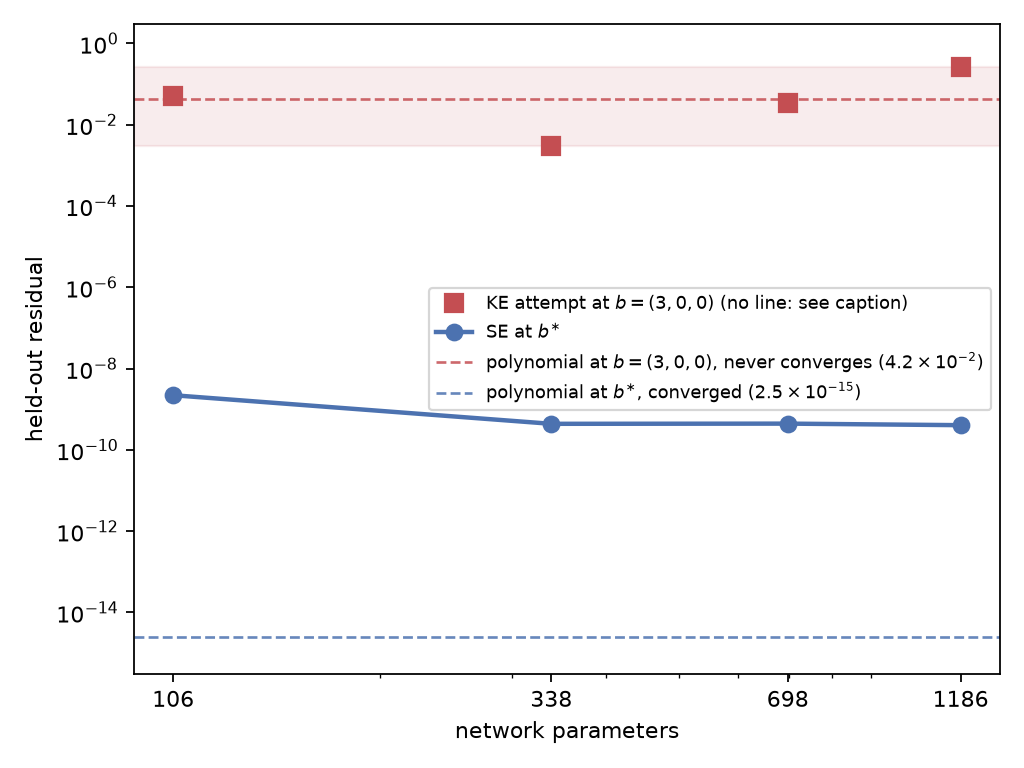}
\caption{Escalating the width on the $\dP2$ pentagon at a released budget,
$10^{5}$ function and gradient evaluations, held-out residual against parameter
count on logarithmic axes. Squares are the K\"ahler--Einstein attempt at the
regular Reeb vector, drawn without a connecting line because the sequence is
not monotone, and the band is the range they occupy. Circles are the same
network on the same pentagon at the volume-minimizing $b^{\ast}$, which
saturates. The dashed lines are the polynomial, labelled by how its optimizer
exits: at $b^{\ast}$ on its own convergence test, at the regular Reeb vector
by exhausting its evaluation budget.}
\label{fig:widthladder}
\end{figure}

We run the same construction at the regular Reeb vector with $\psi$
carried by a network, where the parameter that enlarges the class is its width
rather than a degree. The budget has to be released first: at
\texttt{scipy}'s default limit of $15{,}000$ evaluations every point of such a
sequence stops early and at the same place, so a nearly constant residual across these runs does not establish
convergence with width, and appendix~\ref{app:robust} measures what the released budget is
worth. At $10^{5}$ evaluations the four widths, from $106$ to $1186$
parameters, give held-out residuals of $5.0\times10^{-2}$,
$3.0\times10^{-3}$, $3.4\times10^{-2}$ and $2.7\times10^{-1}$. The same
network on the same pentagon at $b^{\ast}$, at the same budget, gives
$2.2\times10^{-9}$, $4.4\times10^{-10}$, $4.4\times10^{-10}$ and
$4.1\times10^{-10}$ (figure~\ref{fig:widthladder}).

At $b^{\ast}$ the held-out residual saturates with width from $338$ parameters
onward, and the in-sample and held-out values agree within eleven percent. At
the regular Reeb vector both are non-monotone, the held-out residual varying by
a factor $90$ and largest at the greatest width, with in-sample to held-out
ratios of $2.6$--$25$. The separation from the matched-width $b^{\ast}$ runs is
$6.8$--$8.8$ orders of magnitude, against about twelve for the polynomial.
For nested approximation spaces the smallest residual attainable on a fixed
sample cannot increase with width, so the non-monotone in-sample sequence is
not that sequence and the difference between them is optimization error; the
held-out values carry finite-sample effects in addition. How much of the
plateau remains once these are removed, and whether a larger budget would
remove more of it, are not settled here.

The failure signatures differ in kind between the two classes. The polynomial
overfits, its in-sample residual pulling away from the held-out one, which
stays where it is, whereas the network's in-sample residual does not run away;
its gap to the held-out residual is there at every width and does not open as
the width grows. The overfitting
fingerprint is therefore specific to the polynomial. The iteration count
discriminates for the polynomial but not for the network on this pentagon,
where it runs to its cap at $b^{\ast}$ as well --- although the
$D_6$-informed and symmetry-free $\dP3$ networks of
section~\ref{sec:nnpos} both stop on their own convergence test when carried
past their production caps.

\section{The primitive harmonic \texorpdfstring{$(2,1)$}{(2,1)}-forms}
\label{sec:oneone}

The metric and harmonic forms are inputs to non-conformal holographic
solutions in IIB supergravity. On the cone, the three-form flux is specified by a primitive
imaginary self-dual harmonic $(2,1)$-form, $G_3=C\,\omega_{2,1}$, the condition
for unbroken supersymmetry in a warped background
\cite{Grana:2000jj,Giddings:2001yu}, and the warp factor\footnote{The warp factor $h$
appears in $ds^2_{10} = h^{-1/2}ds^2_4 + h^{1/2} ds^2_{X}$, where $ds^2_4$ is the four-dimensional Minkowski metric and $X$ is the six-dimensional Ricci-flat K\"ahler space that carries the flux $G_3$.}
obeys the Poisson equation
\begin{equation}
\triangle_{X} h=-\tfrac{g_s^2}{12}\,
G_{\mu_1\mu_2\mu_3}(G^{\ast})^{\mu_1\mu_2\mu_3} .
\label{eq:warp}
\end{equation}
On the conifold this setup leads to the Klebanov--Tseytlin and
Klebanov--Strassler solutions \cite{Klebanov:2000nc,Klebanov:2000hb}, and on
the cones over $\Ypq$, the cone over $\dP1$ among them, Herzog, Ejaz and
Klebanov (HEK) built the harmonic $(2,1)$-form from the GMSW metric and
determined the warp factor explicitly \cite{Herzog:2004tr}. For the del Pezzo cones
beyond $\dP1$ the dual cascades and their fractional branes have been analysed
on the field-theory side \cite{Franco:2005zu,Bertolini:2005di}; on $\dP1$
itself the HEK solution was used to argue that the cascade
ends in supersymmetry breaking, the complex deformation of the cone being
obstructed \cite{Berenstein:2005xa}. For $\dP3$ DHHKW supplied the first
two inputs of the construction, the K\"ahler--Einstein metric and a numerical
$(1,1)$-form $\theta$ on it, and noted that with these in hand the warp factor
could be computed; to our knowledge that computation has not been carried out
since, and for $\dP2$ neither input existed. Those two inputs are what this
paper supplies on both cones: the metric in the sections above, and
$\omega_{2,1}$ here.

For a regular Sasaki--Einstein link,
\(\omega_{2,1}=(-i\,dr/r+\eta)\wedge\theta\), where \(\theta\) is a primitive
anti-self-dual harmonic \((1,1)\)-form on the transverse K\"ahler--Einstein
surface \cite{Doran:2007zn}. The same construction applies to an irregular
link when \(\theta\) is interpreted as a basic form on \(Y\). Closedness
and primitivity of \(\theta\) give
\(d\omega_{2,1}=d\eta\wedge\theta=0\), since \(d\eta\) is a constant multiple
of the transverse K\"ahler form, and imaginary self-duality of
\(\omega_{2,1}\) is equivalent to transverse anti-self-duality of \(\theta\).
We represent \(\theta\) by a scalar \emph{stream function}
$\chi$ on the polygon. This enforces closedness and primitivity identically,
leaving a linear second-order equation for the $(1,1)$ condition. This condition
is equivalent to anti-self-duality, which we enforce by minimizing the
self-dual energy \eqref{eq:sdenergy}.
Being a statement about \(L^2\) norms it needs a trial space in
\(L^2\), which unconstrained polynomial trial spaces do not provide; the admissible space that does is
rational, and retains the \(d-3\) primitive cohomology directions, where \(d\)
is the number of polygon edges. Minimizing the energy over it is a generalized
eigenvalue problem, and the \(d-3\)
eigenvalues that collapse belong to the forms sought. We test this
construction on the conifold and \(Y^{3,2}\) and compute the representatives
on \(\dP3\) and \(\dP2\).

\subsection{The stream function and the type equation}
\label{sec:oneone-reduce}

We seek torus-invariant two-forms \(\theta\) on the transverse geometry of
section~\ref{sec:setupgeom} that are closed, of type \((1,1)\), and primitive:
\(\theta\wedge\omega_T=0\). At a regular Reeb vector these are forms on the
transverse K\"ahler--Einstein surface. In general, and in particular at an
irregular Reeb vector, they are defined instead as \emph{basic} forms on the
link, annihilated by \(\xi\) and invariant under its flow. The exterior
derivative preserves basic forms, since
\(\iota_\xi d\alpha=\mathcal L_\xi\alpha-d\iota_\xi\alpha=0\); its restriction
\(d_B\) defines the basic cohomology, the cohomology of this subcomplex, and
a basic form is basic-exact when it is \(d_B\) of a basic form. This is
stronger than exactness on \(Y\): \(\omega_T\) is a constant multiple of
\(d\eta\), but \(\eta\) is not basic, and \([\omega_T]\neq0\) in basic
cohomology. At a regular Reeb vector the two descriptions agree, a basic form being the pullback of a form
on the quotient, a smooth surface; for a quasi-regular Reeb vector the
quotient is an orbifold, and for an irregular one the leaf space is not a
manifold, so the basic description is the general one. The transverse
metric \(g_T=g_Y-\eta\otimes\eta\) and orientation
\(\omega_T\wedge\omega_T\) define a Hodge star \(\star_T\) on basic forms.

The computation uses basic tensors on the link \(Y\), reduced by the torus action to
the polygon \(P\). In transverse dimension four a primitive real
\((1,1)\)-form is anti-self-dual, and closedness then gives
\(d\star_T\theta=-d\theta=0\), hence harmonicity. The number of solutions
is fixed by the polygon. Where the Reeb vector is regular the transverse
geometry is a toric surface built from the $d$-gon, with $b_2=d-2$
\cite[\S3.4]{Fulton:1993toric} and $b_2^+=1$, and Hodge theory counts the
solutions as the primitive part of $H^2$,
\begin{equation}
b_2^-=d-3 .
\label{eq:b2minus}
\end{equation}
At an irregular Reeb vector, we consider instead the primitive part of basic cohomology, whose dimension is again
$d-3$. Appendix~\ref{app:admissibility} proves this directly: the boundary values of
the stream function identify primitive basic classes with the $d$ vertex
values modulo the three affine directions, and every such class has a unique
harmonic completion. This argument uses basic Hodge theory but no prior
cohomological dimension count, and applies equally to regular, quasi-regular
and irregular Reeb vectors. The count agrees with a theorem of Goertsches,
Nozawa and T\"oben on K-contact manifolds:\footnote{A contact form $\eta$ with
Reeb vector $\xi$ and a metric $g$ form a K-contact structure when $\xi$ is a
Killing field of $g$ and $g$ is compatible with $d\eta$ through an almost
complex structure on $\ker\eta$ \cite[Defs~2.1--2.2]{GoertschesNozawaToben:2012};
a Sasakian structure is a K-contact structure whose transverse almost complex
structure is integrable, so every metric of this paper is K-contact. A contact
toric manifold is of Reeb type when the Reeb vector of some contact form in the
structure generates a subaction of the torus
\cite[Def.~2.6]{GoertschesNozawaToben:2012}; here $\xi=\sum_Ib_I\partial_{\phi_I}$,
so every link of this paper is of Reeb type.} for a compact contact toric
manifold of Reeb type, the basic cohomology of the Reeb foliation has total
dimension equal to the number of vertices of its moment polytope
\cite[Thm~7.11 and Example~9.23]{GoertschesNozawaToben:2012}, here $d$. The
$d-3$ primitive classes, the class of $\omega_T$, and the classes of $1$ and
of $\omega_T\wedge\omega_T$ in degrees $0$ and $4$ exhaust that dimension. We
use $b_2^-$ for a quotient surface and the primitive basic dimension otherwise.

The frame in which this is written has to be chosen with the irregular case in
mind as well, since there the three angles $\phi_I$ of section~\ref{sec:setupgeom} do
not split into a Reeb angle and two transverse ones. What does exist for every
$b$ is the pair of constant-coefficient one-forms
\begin{equation}
\nu_i\equiv\langle e_i,d\phi\rangle= (e_i)_I\,d\phi_I ,\qquad i=1,2,
\label{eq:nui}
\end{equation}
built from the same orthonormal basis $e_{1,2}$ of $\ker b$ that defines the
slice chart $s$ in \eqref{eq:slicechart}. They annihilate the Reeb vector,
$\nu_i(\xi)=\langle e_i,b\rangle=0$, and are invariant along it, so they
are basic, though not by themselves smooth forms on $Y$: at a facet the circle
generated by $v_a$ collapses, and only combinations whose coefficients vanish
there in the right way extend, as appendix~\ref{app:admissibility} makes precise. At a rational $b$ they are, up to normalization, the differentials
of angles on the quotient, but nothing below uses that.

In the symplectic frame
$\omega=\sum_I dy^I\wedge d\phi_I$. Inverting $t=y/\ell_b(y)$ of
section~\ref{sec:param} writes a point of the moment cone as $y=\ell_b\,t$
with $t=t_0+s_ie_i$ from \eqref{eq:slicechart}, so that $(\ell_b,s_1,s_2)$ is a
chart on its interior and $dy=t\,d\ell_b+\ell_b\,e_i\,ds_i$. Pulled back to
the level set $\ell_b=1$, where $d\ell_b=0$ and $dy=e_i\,ds_i$, the symplectic
form is $\sum_i ds_i\wedge\langle e_i,d\phi\rangle=\sum_i ds_i\wedge\nu_i$, a
basic form on the transverse geometry. In this section we use the slice
normalization $\omega_T=\sum_i ds_i\wedge\nu_i$, twice the transverse
K\"ahler form of the link at $r=1$ in section~\ref{sec:graders}.
This is also the normalization of appendix~\ref{app:conventions}; the
conditions below are independent of this factor: primitivity, the type condition, the Hodge star on two-forms and the
energy ratio are unchanged by a constant rescaling. The corresponding slice metric is
$\tilde g_T=2g_T=(\Hess G_P)_{ij}\,ds_i\,ds_j+u^{ij}\,\nu_i\nu_j$,
with $u=(\Hess G_P)^{-1}$
as in section~\ref{sec:graders}, since $\partial_i\partial_jG_P=(e_i)_I(e_j)_JG_{IJ}$
on the slice.

A torus-invariant basic two-form of the mixed type that
$i\partial\bar\partial$ of an invariant function produces is then
\begin{equation}
\theta=A_{ij}\,ds_i\wedge\nu_j ,
\label{eq:thetaA}
\end{equation}
with $A$ a matrix-valued function on the polygon. The following discussion
allows us to parametrize $A$ by a single scalar. The mixed ansatz captures every torus-invariant anti-self-dual harmonic
two-form: a general
torus-invariant basic two-form adds $B(s)\,ds_1\wedge ds_2$ and
$\Gamma(s)\,\nu_1\wedge\nu_2$, closedness makes $\Gamma$ constant, smoothness
across a collapsing circle then forces $\Gamma=0$, and anti-self-duality forces
$B=0$. Closedness implies $A_{ij}=\partial_i\mathbf a_j$ for some vector potential
$\mathbf a$ on the polygon, and
primitivity, $\theta\wedge\omega_T=0$, is $\mathrm{tr}\,A=\mathrm{div}\,\mathbf a=0$.
In two dimensions a
divergence-free field is a rotated gradient,
$\mathbf a=(\partial_2\chi,-\partial_1\chi)$, so both conditions are solved
identically by a single \emph{stream function} $\chi$ on the polygon,
\begin{equation}
A=\Hess(\chi)\begin{pmatrix}0&-1\\[2pt]1&0\end{pmatrix}
=\begin{pmatrix}\chi_{12}&-\chi_{11}\\[2pt]\chi_{22}&-\chi_{12}\end{pmatrix},
\qquad \chi_{ij}:=\partial_i\partial_j\chi ,
\label{eq:streampot}
\end{equation}
with $\chi$ unique up to an affine function; that is the
gauge the count below quotients by. The one condition left, type $(1,1)$, is a
single linear second-order equation for $\chi$,
\begin{equation}
u^{ij}\,\partial_i\partial_j\chi=0\qquad\text{in the interior of }P,
\label{eq:oneoneeq}
\end{equation}
with $u^{ij}$ the inverse slice Hessian of section~\ref{sec:graders}. Note that it is
not the same as the transverse Laplacian $\triangle=\partial_i(u^{ij}\partial_j\,\cdot)$: $u$ is not differentiated in (\ref{eq:oneoneeq}), and the two operators differ by the
first-order term $(\partial_iu^{ij})\partial_j$.

DHHKW also use a scalar reduction, following Abreu
\cite{Abreu:1998xhs,Abreu:2000xhs}. One may write
$\theta_a=i\partial\bar\partial\mu_a$, in our normalization, with a potential
$\mu_a$ that has a logarithmic singularity at the boundary. Closedness is then
automatic, and
harmonicity becomes the constancy of $\triangle\mu_a$
\cite[(3.46)]{Doran:2007zn}, primitivity being the vanishing of that constant
\cite[(3.48)]{Doran:2007zn}. The difference between their reduction and
\eqref{eq:streampot} matters numerically: in theirs primitivity is a condition
to be imposed, and therefore a source of error, while in \eqref{eq:streampot}
it is an identity of the ansatz. In
section~\ref{sec:dp3grading} we have graded DHHKW's published truncation on our
metrics; here we solve for the representative in the parametrization
\eqref{eq:streampot}, which builds in primitivity where theirs builds in the
type condition.

Near facet $a$, let us choose a chart where the facet is $x=0$ with tangential
coordinate $z$. The slice Hessian for coordinates $(x,z)$ is then
$\Hess G_P\simeq\mathrm{diag}((2x)^{-1},g_{zz})$ with $g_{zz}>0$, and
\eqref{eq:oneoneeq}, multiplied by the finite positive $g_{zz}$, reads up to
terms subleading at the facet
\[
2x\,g_{zz}\,\chi_{xx}+\chi_{zz}=0 .
\]
The coefficient of the normal second derivative vanishes linearly at the facet
while that of the tangential one does not, and that decides the boundary
behaviour. Write $\chi(x,z)=\chi_0(z)+\dots$ and solve for the next term: a trace
with $\chi_0''(z)\neq0$ forces
\[
\chi=\chi_0(z)-\frac{\chi_0''(z)}{2g_{zz}}\,x\log x+O(x),
\]
continuous but not $C^1$ at the facet. A solution smooth up to the boundary
therefore has $\chi_{zz}=0$ on every facet: its \emph{boundary trace}, the
restriction $\chi|_{\partial P}$, is affine on each of the $d$ edges, and
neither another Dirichlet datum nor an independent
Neumann datum can be prescribed there. We call the restriction to a single
edge the \emph{edge trace} of $\chi$; the tangential second derivative
$\chi_{zz}$ on that edge measures its departure from an affine function. The general theory of such boundary
degenerations, due to Keldysh, is given in \cite{OleinikRadkevich:1973,Otway:2012};
DHHKW met the same degeneration in their operator \cite[\S3.4]{Doran:2007zn}.

What these traces mean for the classes is settled in appendix~\ref{app:admissibility}.
The class of $\theta_\chi$ in the primitive basic cohomology of the link depends
only on the $d$ vertex values of $\chi$ up to shifts by affine functions, a
three-dimensional freedom that leaves $\theta$ unchanged; every
such class arises in this way, and for each choice of vertex values there is
exactly one $\chi$ with affine edge traces that solves \eqref{eq:oneoneeq},
whose form is the basic harmonic representative of its class. This holds at
regular, quasi-regular and irregular Reeb vectors and needs only that the
metric is a smooth toric Sasakian metric, not that it is Einstein. The solution space of
\eqref{eq:oneoneeq} modulo affine functions therefore has dimension $d-3$,
which recovers \eqref{eq:b2minus} without assuming it.
Figure~\ref{fig:streampot} shows the stream functions of all $d-3$
representatives on the two polygons.

\subsection{The self-dual energy and the admissible trial space}
\label{sec:oneone-energy}

We use a ratio of the form \eqref{eq:rayleigh} built on the self-dual
part of $\theta$, which vanishes exactly on an anti-self-dual form,
\begin{equation}
\mathcal E[\theta]=\frac{\int_P\bigl\|\tfrac12(\theta+\star_T\theta)\bigr\|^2}
                {\int_P\|\theta\|^2}\;\in\;[0,1] .
\label{eq:sdenergy}
\end{equation}
The interval $[0,1]$ is the general bound for the orthogonal projection
$\tfrac12(1+\star_T)$. In the closed primitive trial space constructed below,
the sharper bound $\mathcal E\le\tfrac12$ holds, with equality precisely
for nonzero basic-exact forms, as the argument leading to
\eqref{eq:distance} shows.

Expressed through $\chi$, the numerator density is
$\|\theta_+\|^2=\tfrac12\det \Hess G_P\,(u^{ij}\chi_{ij})^2$, in
the flat measure of the polygon, which is the Riemannian one since
$\det \tilde g_T=1$. So $\mathcal E$ is a weighted mean square of the residual of
\eqref{eq:oneoneeq}, normalized by $\|\theta\|_{L^2}^2$, and $\mathcal E=0$ is
exactly the type condition: its minimization is a least-squares solution of
\eqref{eq:oneoneeq} in the $L^2$ geometry of the forms. In any finite trial
space closedness and primitivity are exact, and the type condition holds to
the extent that $\mathcal E$ is small; so does the imaginary self-duality of
the cone form $\omega_{2,1}$, which it implies. The
minimization reduces to a generalized eigenvalue problem
and requires no nonlinear optimization. For nested admissible trial spaces,
the Ritz eigenvalues cannot increase as the space is enlarged. A small value
therefore exhibits a nearly anti-self-dual form in $L^2$, whereas a large
value may simply reflect an insufficient trial space. For that argument every
trial form has to be in $L^2$ to begin with, which is not automatic for a
polynomial $\chi$ and is arranged next.

\label{sec:oneone-adm}
In the facet chart
$(x,z)$ of section~\ref{sec:oneone-reduce}, the pointwise norm of
\eqref{eq:thetaA} is to leading order at the facet
\begin{equation}
\|\theta\|^2=\mathrm{tr}\bigl(u\,A\,\Hess G_P\,A^{\mathrm{T}}\bigr)
\simeq2\chi_{xz}^2+2x\,g_{zz}\,\chi_{xx}^2+\frac{\chi_{zz}^2}{2x\,g_{zz}}\,.
\label{eq:facetnorm}
\end{equation}
For a trial function smooth up to the boundary, a nonzero
$\chi_{zz}(0,z)$ makes the last term logarithmically non-integrable,
so $\theta\notin L^2$. The same boundary defect appears in the type
residual, the left-hand side of \eqref{eq:oneoneeq}. Since the second
derivatives of $\chi$ are bounded, the inverse-Hessian asymptotics of
section~\ref{sec:oneone-reduce} give, away from the vertices,
\[
u^{ij}\chi_{ij}
=\frac{\chi_{zz}(0,z)}{g_{zz}(0,z)}+O(x).
\]
Thus the residual does not vanish at the edge unless the edge trace of
$\chi$ is affine. The determinant asymptotics then give the numerator density:
\[
\det\Hess G_P=\frac{g_{zz}(0,z)}{2x}+O(1),\qquad
\|\theta_+\|^2=\frac{\chi_{zz}(0,z)^2}{4x\,g_{zz}(0,z)}+O(1).
\]
Both integrals defining \eqref{eq:sdenergy} therefore diverge for a smooth
trial function with a non-affine edge trace. This argument uses bounded
second derivatives: for the non-smooth $x\log x$ solution discussed in
section~\ref{sec:oneone-reduce}, the singular normal derivative can instead
cancel the tangential contribution to the type residual.
The exact smooth solution has $\chi$
affine along every edge, as section~\ref{sec:oneone-reduce} derived, but a
trial polynomial does not, and Gauss--Legendre nodes never reach the edge, so a
finite quadrature value says nothing by itself about integrability. On the
hexagon, pentagon and $Y^{3,2}$ quadrilateral, globally polynomial stream
functions cannot provide an admissible space containing all primitive
classes: non-affine edge traces produce forms outside $L^2$, as
\eqref{eq:facetnorm} shows, while exactly affine edge traces leave only
globally affine vertex data, as appendix~\ref{app:admissibility} proves.

We therefore use a rational admissible trial space. Let $V_1,\ldots,V_d$ be
the vertices of $P$, numbered so that edge $a$, the one on the line
$\ell_a=0$, runs from $V_{a-1}$ to $V_a$, and let
$\varphi_a$ be the Wachspress coordinates of the convex polygon
\cite{Wachspress:1975}: rational functions, smooth on the closed polygon, with
$\varphi_a(V_b)=\delta_{ab}$, exactly affine along every edge, and satisfying
$\sum_a\varphi_a=1$ and $\sum_aV_a\varphi_a=s$. In the second identity $s$ is
the point of $P$ itself, so the two together make the $\varphi_a$ generalized
barycentric coordinates, writing each $s\in P$ as the convex combination
$\sum_a\varphi_a(s)\,V_a$ of the vertices; equivalently, they reproduce affine
functions, since $f(s)=\alpha+\langle\beta,s\rangle$ has
$\sum_af(V_a)\varphi_a(s)=\alpha\sum_a\varphi_a(s)
+\bigl\langle\beta,\sum_aV_a\varphi_a(s)\bigr\rangle=f(s)$. We expand
\begin{equation}
\chi=\sum_a c_a\,\varphi_a+\prod_a\ell_a\cdot q(s)\,,\qquad \deg q\le N-d\,,
\label{eq:admspace}
\end{equation}
where $q(s)$ is an arbitrary polynomial, free of any condition of its own because
the prefactor already vanishes on $\partial P$; the coefficients $c_a$
together with those of $q(s)$ are the parameters. Affine reproduction makes the
vectors $c_a=f(V_a)$ a three-dimensional subspace of $\mathbb R^d$, the
vertices affinely spanning the plane. With $q=0$, these coefficients give
an affine $\chi$ and hence $\Hess\chi=0$. We quotient out these three
Wachspress combinations and call the $d-3$ surviving ones the \emph{vertex
directions}, and the functions $\prod_a\ell_a\cdot q$ the \emph{bubble
directions}. That count is the one of section~\ref{sec:oneone-reduce}, read
off the ansatz.

Every member has affine edge traces, so every trial form is in $L^2$ and the
type residual vanishes at the facets. The spaces increase with $N$, and their
union is dense in the admissible class $\mathcal A(P)$ of
appendix~\ref{app:admissibility} modulo affine functions, with respect to the
$C^2(\overline P)$ quotient norm. Indeed, a $\chi$ in that class with vertex values
$c_a$ agrees with $\sum_ac_a\varphi_a$ along every edge, two affine functions
on a segment with equal endpoint values being equal, so it differs from it by
$\prod_a\ell_a$ times a smooth function, which polynomials approximate
together with its second derivatives, the derivatives that enter $\theta$.
The Wachspress part is rational and fixed independently of $N$, while the
bubble part is a polynomial of total degree at most $N$, with the product
of $d$ affine factors contributing degree $d$. Thus $N$, which we call the 
\emph{order of the enrichment}, measures the trial functions on the same scale as the
degrees of the polynomial fits of the metric in sections~\ref{sec:dp3}
and~\ref{sec:dp2}, and it is the parameter varied in the convergence tests below.
Every number reported for our own forms is computed in \eqref{eq:admspace};
the energies quoted for DHHKW's forms in section~\ref{sec:oneone-dp3} use
their span.

Two identities determine the interpretation of the eigenvalues. Write
$M_{ij}=\langle\theta_i,\theta_j\rangle$, $T_{ij}=\int\theta_i\wedge\theta_j$
and $S_{ij}=\langle(\theta_i)_+,(\theta_j)_+\rangle$ on a trial basis, $T$
being the intersection pairing on that basis.
The pointwise identity
$\|\theta_+\|^2=\tfrac12\bigl(\|\theta\|^2+\theta\wedge\theta/dV\bigr)$,
integrated over pairs of trial forms, gives $S=\tfrac12(M+T)$. By appendix~\ref{app:admissibility}, $T$ vanishes on the bubble directions and is negative definite
on the vertex directions, so $T$ has rank $d-3$ with negative nonzero
eigenvalues; the eigenvalue problem for $\mathcal E$ therefore has all but
$d-3$ of its eigenvalues at exactly $\tfrac12$, for any compatible K\"ahler
metric and each enrichment order. Neither that count nor the value $\tfrac12$
is a numerical result. The computation determines the magnitudes of
the $d-3$ energies below it and their decrease with $N$. The second identity
relates $\mathcal E$ to an $L^2$ distance. Where a
surface exists $b_2^+=1$, so the harmonic representative $\theta_{\mathrm{harm}}$ of the class of a
closed primitive trial $\theta$ is primitive and anti-self-dual; then
$\theta=\theta_{\mathrm{harm}}+d\alpha$ with $\langle \theta_{\mathrm{harm}},d\alpha\rangle=0$ and
$\int\theta\wedge\theta=\int \theta_{\mathrm{harm}}\wedge \theta_{\mathrm{harm}}=-\|\theta_{\mathrm{harm}}\|^2$, whence
$\mathcal E=\tfrac12\bigl(1-\|\theta_{\mathrm{harm}}\|^2/\|\theta\|^2\bigr)$ and
\begin{equation}
\frac{\|\theta-\theta_{\mathrm{harm}}\|_{L^2}}{\|\theta\|_{L^2}}=\sqrt{2\,\mathcal E}\,.
\label{eq:distance}
\end{equation}
At an irregular Reeb vector the same holds with the basic Hodge decomposition
in place of the surface one: the primitive basic harmonic representative is
anti-self-dual and basic-exact directions drop out of the pairing
(appendix~\ref{app:admissibility}). For the numerical metric held fixed,
a nonzero distance measures the error in approximating its harmonic
representative. The effect of error in the metric itself is not included.

The integrals use the deterministic centroid-fan rule of
section~\ref{sec:graders} over the whole polygon, with no excluded margin.
We orthonormalize the trial forms in $L^2$ by rank-revealing pivoted QR and
solve the generalized eigenproblem. The reported energies are then evaluated
directly from the fixed stream functions. They are stable under quadrature
refinement and alternative evaluations of the type residual
(appendix~\ref{app:robust}, table~\ref{tab:quadscan}). These checks assess
numerical evaluation, not the error in the computed vectors; the smallest
conifold and $Y^{3,2}$ energies are at the arithmetic floor.

\subsection{Divisor periods and the choice of basis}
\label{sec:oneone-periods}

When the transverse geometry is a smooth quotient surface, its Picard
lattice provides the integral classes used to normalize the harmonic forms.
For $\dP n$ with $n=2,3$, we use the basis consisting of the hyperplane
class $H$ and the exceptional classes $E_i$
\cite{Manin:1974cubic,Dolgachev:2012classical}. Their intersection pairing
and the canonical class $K$ are given by
\[
 H^2=1,\qquad E_i\cdot E_j=-\delta_{ij},\qquad H\cdot E_i=0,
 \qquad K=-3H+\sum_iE_i.
\]
Since the K\"ahler class is anticanonical, $[\omega_T]\propto-K$, primitive
classes are orthogonal to $K$; on the toric surface $-K=\sum_aD_a$, where
$D_a$ is the toric divisor over edge $a$, the surface on which the circle
generated by $v_a$ collapses. On $\dP3$,
$K^\perp=A_2\oplus A_1$ has root basis
$\alpha_1=E_1-E_2$, $\alpha_2=E_2-E_3$ and $\beta=H-E_1-E_2-E_3$,
with Gram matrix
\[
 \begin{pmatrix}-2&1&0\\1&-2&0\\0&0&-2\end{pmatrix}.
\]
The six $D_a$ are the $(-1)$-curves of $\dP3$, the rational curves of
self-intersection $-1$: the exceptional curves $E_1,E_2,E_3$ and the lines
$L_{ij}=H-E_i-E_j$ through pairs of blown-up points, each meeting its two
neighbours once.
Their cyclic permutation by the order-six rotation acts on the $A_2$
block with eigenvalues $e^{\pm2\pi i/3}$ and on $\beta$ with eigenvalue
$-1$, fixing the splitting used below.

On $\dP2$, the basis $E_1-E_2$, $H-E_1-2E_2$ of $K^\perp$ has Gram $\bigl(\begin{smallmatrix}-2&-1\\-1&-4\end{smallmatrix}\bigr)$,
with determinant $7$; its only roots are $\pm(E_1-E_2)$, so roots do not
span this rank-two lattice. This surface lattice is not identified with an
integral lattice of basic classes at the irregular $b^\ast$.

The following divisor-period construction applies to the regular hexagon.
Its divisor periods fix the cohomology classes and normalization of the
harmonic two-forms $\theta$. Each period is computed from the rotated gradient
$\mathbf a=(\partial_2\chi,-\partial_1\chi)$ at the two ends of an edge,
without quadrature. Here $b=(3,0,0)$, so the slice frame identifies the
angular lattice modulo the Reeb direction with the projected lattice. With
the edge endpoints $V_{a-1},V_a$ numbered as in section~\ref{sec:oneone-adm},
let $m_a$ be a vector of that lattice which, modulo the collapsing circle,
represents a primitive generator of the circle surviving on $D_a$. Then
\begin{equation}
P_a\;=\;\frac1{2\pi}\int_{D_a}\theta
\;=\;\bigl\langle m_a,\;\mathbf a(V_a)-\mathbf a(V_{a-1})\bigr\rangle .
\label{eq:period}
\end{equation}
The formula follows by restricting $\theta=d(\mathbf a_j\nu_j)$ to
$D_a$: integration around the surviving circle gives $2\pi$, while the
remaining integral is a total derivative along the edge. The circle
collapses at both endpoints, making $D_a$ a two-sphere.

The \emph{lattice length} $L_a$ of an edge is defined by
\[
V_a-V_{a-1}=L_a\,t_a,\qquad L_a>0,
\]
where $t_a$ is the primitive vector of the moment lattice directed along
the edge. Thus the Euclidean length is $L_a|t_a|$, which need not equal
$L_a$. We orient $m_a$ so that $\langle m_a,t_a\rangle=1$, giving
$\langle m_a,V_a-V_{a-1}\rangle=L_a$. For the hexagon of
figure~\ref{fig:polytopes}, the vertices in the display chart $\tilde s=3s$ are the
six lattice points $(\pm1,0)$, $(0,\pm1)$ and $\pm(1,-1)$, so every edge
vector is a primitive lattice vector: $L_a=1$ in $\tilde s$, hence $L_a=1/3$ on
the slice $\ell_b=1$, while the Euclidean edge lengths in $\tilde s$ are $1$ for
four edges and $\sqrt2$ for two. As a normalization check,
$\mathbf a=s$ gives the slice symplectic form and \eqref{eq:period} returns
$1/3$. The reduced periods $P_a$ are integers when
$[\theta/(2\pi)]$ is an integral class.

The choice of $m_a$ modulo the collapsing circle does not affect the result.
Writing $n_{a,j}=\langle v_a,e_j\rangle$, the change
$m_a\mapsto m_a+c\,n_a$ leaves \eqref{eq:period} unchanged because the affine
edge trace makes $\langle n_a,\mathbf a\rangle$ constant along the edge.

Primitivity implies $\sum_aP_a=0$, since $-K=\sum_aD_a$. On the hexagon
the three representatives are fixed by prescribing their periods to be those
of $\alpha_1$, $\alpha_2$ and $\beta$ (section~\ref{sec:oneone-dp3}); on the
pentagon, where the irregular Reeb vector leaves no quotient surface, the
basis is fixed by the lattice involution of the pentagon instead
(section~\ref{sec:oneone-dp2}).

\subsection{Closed-form references: the conifold and \texorpdfstring{$Y^{3,2}$}{Y(3,2)}}
\label{sec:oneone-closed}

The conifold and $Y^{p,q}$ come with closed-form harmonic forms, and they
calibrate the construction before it is applied where no reference exists. On
the conifold the exact stream function is a polynomial that the trial space
contains, so the check is exact. For $Y^{p,q}$ we do not compare $\theta$ or
$\chi$ pointwise: the closed form is written in coordinates and a frame
adapted to that geometry, and matching them to the slice chart would
introduce a coordinate bridge of its own. Instead we compare a quantity that
is free of normalization and frame, the ratio of the extreme amplitudes of
the form over the polygon, and check that the count of collapsing energies is
the expected one.

On the conifold, which has one primitive direction by \eqref{eq:b2minus},
the exact stream function is the quadratic $\chi=uv$ in affine coordinates
aligned with two adjacent edges (appendix~\ref{app:admissibility}). It lies
in the trial space from $N=4$ on, and the lowest self-dual energy is at the
arithmetic floor at every tested order and under both quadrature rules.

For the $Y^{p,q}$ family, HEK's closed-form primitive $(1,1)$-form is
\cite[(29)--(30)]{Herzog:2004tr}
\[
\theta=F(y)\bigl(e^1\wedge e^2-e^3\wedge e^4\bigr),
\qquad F(y)=(1-y)^{-2},
\]
where $e^1,\ldots,e^4$ form an orthonormal coframe for the four-dimensional
transverse K\"ahler metric. We choose $p=3$, $q=2$ to test the construction
at an irregular Reeb vector. The ratio of the maximum and minimum
amplitudes is independent of the normalization:
\begin{equation}
R=\frac{F_{\max}}{F_{\min}}
=\Bigl(\frac{1-y_1}{1-y_2}\Bigr)^2
=7+2\sqrt6\simeq11.898979486.
\label{eq:hekratio}
\end{equation}
Here $y_1=-1/\sqrt6$ and $y_2=1-1/\sqrt6$ are the GMSW endpoints
(appendix~\ref{app:ypq}). For the computed form, we evaluate the same ratio
using the positive eigenvalue $\lambda=\sqrt{-\det\Hess\chi}$ of $A$; the
passage to an orthonormal transverse frame is a similarity, so for the exact
solution $\lambda$ is proportional to $F$, with a constant that cancels in
the ratio.

The computation uses the degree-$12$ orthonormal polynomial metric fit of
section~\ref{sec:whiten} (table~\ref{tab:sweep}), so the quoted errors include
the effects of the metric approximation. With two trial functions at
$N=4$, the relative error in $R$ is $1.4\times10^{-8}$; increasing the form
trial degree to $N=16$ reduces it to $9.9\times10^{-10}$.
At $N=16$ there is one small self-dual energy, $4\times10^{-18}$ at the
arithmetic floor, with the remaining eigenvalues at $\tfrac12$, as expected
for $d-3=1$.

\subsection{The hexagon: three representatives on \texorpdfstring{$\dP3$}{dP3}}
\label{sec:oneone-dp3}

For $\dP3$, $b_2^-=3$, so we expect three small self-dual energy
eigenvalues approaching zero as the trial space is enriched, with all
remaining eigenvalues equal to $\tfrac12$
(section~\ref{sec:oneone-adm}).
The computation on our degree-$18$ metric confirms this pattern: at
$N=16$, with $69$ trial functions, the three energies are of order
$10^{-9}$, and at $N=20$, with $123$, of order $10^{-12}$--$10^{-11}$, two
of them degenerate, while the remaining eigenvalues stay at
$\tfrac12$ (the values are in table~\ref{tab:quadscan}). The double degeneracy follows from the
$D_6$ symmetry of the metric and trial space, which acts irreducibly on
the two-dimensional $A_2$ component of $K^\perp$.

The stream functions in figure~\ref{fig:streampot} generate the forms
$\theta_\gamma$ through \eqref{eq:streampot}, with
$\gamma\in\{\alpha_1=E_1-E_2,\alpha_2=E_2-E_3,\beta=H-E_1-E_2-E_3\}$.
Their classes and scales are fixed by the divisor periods
\[
\frac{1}{2\pi}\int_{D_a}\theta_\gamma=\gamma\cdot D_a.
\]
Numbering the edges in the fan order of figure~\ref{fig:polytopes}, the
divisors in cyclic order are $D_1,\ldots,D_6=E_1,L_{12},E_2,L_{23},E_3,L_{13}$,
with $L_{ij}=H-E_i-E_j$; any other identification compatible with the
intersection form differs from this one by a symmetry of the hexagon, which
acts on the root basis $(\alpha_1,\alpha_2,\beta)$ by the corresponding
automorphism of $K^\perp$. In that order the reduced periods
are
\[
\begin{array}{c|rrrrrr}
 & E_1 & L_{12} & E_2 & L_{23} & E_3 & L_{13} \\\hline
\theta_{\alpha_1} & -1 & 0 & 1 & -1 & 0 & 1 \\
\theta_{\alpha_2} & 0 & 1 & -1 & 0 & 1 & -1 \\
\theta_\beta & 1 & -1 & 1 & -1 & 1 & -1
\end{array}
\]
Let
$(Q_{\rm int})_{ab}=D_a\cdot D_b$ be the integer intersection matrix of the
six divisors, and let $Q_{\rm int}^{+}$ denote its Moore--Penrose
pseudoinverse. For the primitive fan generators $w_a=(w_a^1,w_a^2)$
of the toric surface, the two divisor relations are
\[
\sum_{a=1}^6 w_a^1[D_a]=0,
\qquad
\sum_{a=1}^6 w_a^2[D_a]=0.
\]
Their coefficient vectors span the kernel of $Q_{\rm int}$, which
therefore has rank $b_2=4$.

The three representatives, whose stream functions at $N=16$ are the top
row of figure~\ref{fig:streampot}, are the combinations of the low-energy
eigenvectors of section~\ref{sec:oneone-adm} whose period vectors are those
of $\alpha_1$, $\alpha_2$ and $\beta$. In the admissible space the periods
are exact linear functions of the Wachspress coefficients, the bubble part
having vanishing gradient at the vertices, so this prescription is a linear
solve; $\sum_aP_a=0$ holds identically for the primitive ansatz, and the
intersection Gram $\mathbf P^{\mathrm{T}}Q_{\rm int}^{+} \mathbf P'$ of the representatives is
the root Gram of $A_2\oplus A_1$ by construction. The Gram is computed
from the periods alone: a class $[\theta/2\pi]=\sum_ac_aD_a$ has period vector
$\mathbf P=Q_{\rm int}c$, so its pairing $c^{\mathrm{T}}Q_{\rm int}c'$ equals
$\mathbf P^{\mathrm{T}}Q_{\rm int}^{+}\mathbf P'$, and the kernel directions of $c$ drop out.
In this identity $\int\theta\wedge\theta'=\mathbf P^{\mathrm{T}}Q_{\rm int}^{+}\mathbf P'$ the
integral is over the transverse geometry with the angular volume $(2\pi)^2$
divided out, the convention of \eqref{eq:period}. Applied
to the symplectic form, whose six periods are $\tfrac13$, both sides equal
$K^2/9=\tfrac23$, twice the area of the hexagon in the chart $s$, which fixes
the normalization of that identity. For an exactly anti-self-dual form
$\int\theta\wedge\theta=-\|\theta\|^2$; at finite $N$ the $L^2$ Gram $M$
differs from minus the intersection Gram by $2S$, $M=-T+2S$ in the notation
of section~\ref{sec:oneone-adm}, so each representative, of
self-intersection $-2$, has $\|\theta\|^2=2/(1-2\mathcal E[\theta])$, which
approaches the common value $2$ in the harmonic limit; the scale shared by
the three panels of figure~\ref{fig:streampot} is that of the intersection
pairing.

DHHKW fitted divisor forms at sixth order \cite[\S6.2]{Doran:2007zn}.
Their logarithmic potentials yield smooth divisor curvature forms, while
the polynomial corrections add smooth exact forms, so the fitted forms are
in $L^2$. We evaluate the self-dual energy on the span of their six fitted
forms, using our degree-$18$ metric and the conventions of
appendix~\ref{app:conventions}. The individual divisor classes are not
primitive; combinations $\sum_a c_a\theta_a$ have primitive classes when
$\sum_a c_a=0$. The three low-energy directions represent these primitive
classes, with energies between $10^{-6}$ and $10^{-5}$. Refitting their
ansatz on our metric by linear least squares on the constancy condition, at
tenth order, gives energies comparable to our $N=16$ results. This comparison concerns finite approximations at different
orders and with different fitting procedures.

\subsection{The pentagon: two basic forms on \texorpdfstring{$\dP2$}{dP2}}
\label{sec:oneone-dp2}

The pentagon is treated by the same admissible space and self-dual energy
minimization of section~\ref{sec:oneone-energy}, at the irregular Reeb vector
$b^{\ast}$ of section~\ref{sec:dp2se} and without the $D_6$ symmetry of the
hexagon. No closed-form or published numerical form exists on this link for
comparison; to our knowledge the two forms below are the first.

On the degree-$14$ metric, two energies lie below
the structural value $\tfrac12$ of section~\ref{sec:oneone-adm}, and they fall
from about $10^{-5}$ at $N=8$ to $10^{-7}$ at $N=12$ and $10^{-10}$ at
$N=16$ (the $N=16$ values are in table~\ref{tab:quadscan}). Their
eigenvectors span the two-dimensional low-energy space, in
which a basis remains to be chosen. At the irregular $b^{\ast}$ the Reeb flow
does not close, so there is no quotient surface and no divisor $D_a$ over
which to integrate \eqref{eq:period}. The fan has exactly two lattice
automorphisms; the nontrivial one leaves $b^{\ast}$ invariant and induces on
the slice an affine involution $s\mapsto f(s)=Rs+\tau$ with $R$ not
orthogonal, the pentagon being affinely but not Euclidean symmetric. It fixes
one vertex and exchanges the remaining four in two pairs, so on the vertex values
modulo affine functions, which are the classes of
appendix~\ref{app:admissibility}, it acts with eigenvalues $+1$ and $-1$, one
each, and its two eigenvectors are determined uniquely up to normalization.

\begin{figure}[tbp]
\centering
\includegraphics[width=0.96\textwidth]{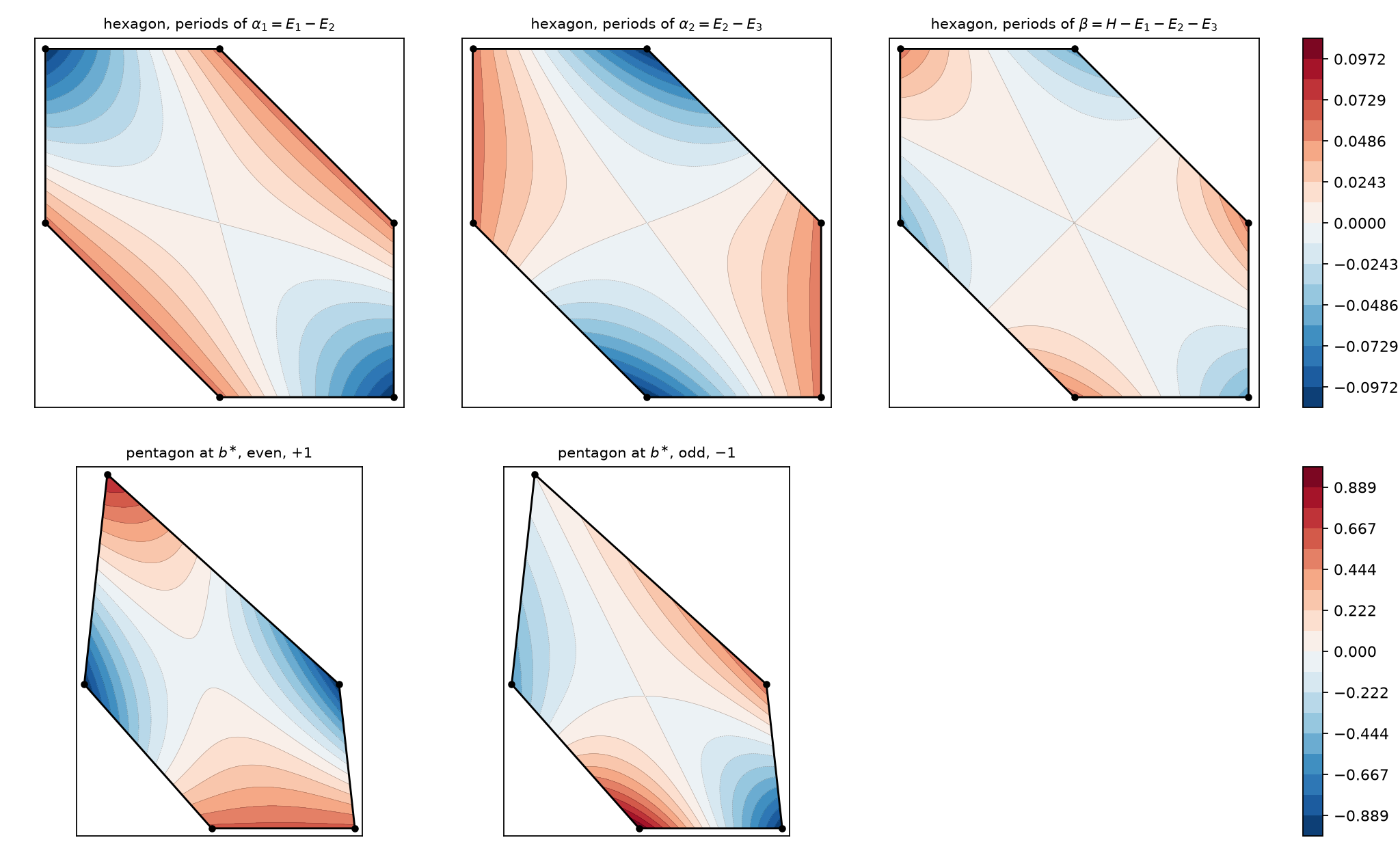}
\caption{Stream functions $\chi$ of all $d-3$ primitive representatives,
computed in the admissible space \eqref{eq:admspace} at $N=16$; the forms are
their rotated Hessians \eqref{eq:streampot}. The affine gauge is that of
\eqref{eq:admspace}: the vertex values of $\chi$ are orthogonal, on the vertex
set, to the constant and the two coordinates, and the bubble part vanishes at
the vertices. Bases are fixed by divisor periods or parity. Top, the hexagon: the stream functions generating $\theta_\gamma$ with
$(2\pi)^{-1}\int_{D_a}\theta_\gamma=\gamma\cdot D_a$ for the three roots
$\gamma=\alpha_1,\alpha_2,\beta$ of $K^{\perp}=A_2\oplus A_1$
(section~\ref{sec:oneone-dp3}). This fixes their scale, so the three share one colour bar and reach $0.109$, $0.109$ and $0.054$.
All three have self-intersection $-2$ and squared $L^2$ norm
$2/(1-2\mathcal E)$, approaching the common harmonic value $2$ as
$\mathcal E\to0$. Bottom, the pentagon at $b^{\ast}$: the even and odd members of its
affine lattice involution, which parity fixes only up to scale, so each panel
is divided by its own $\max|\chi|$ and the two are not to be compared in
amplitude.}
\label{fig:streampot}
\end{figure}

The degree-$14$ metric was fitted without this symmetry and carries it only
approximately: $\Hess G_P$ is invariant to $9.4\times10^{-6}$ relative,
measured as $\max|R^{\mathrm{T}}\Hess G_P(f(s))R-\Hess G_P(s)|$ over the
quadrature nodes against $\max|\Hess G_P|$. The involution therefore preserves
the computed low-energy space only approximately, and we determine its action
on that space by a fit. Let $\chi^{(1)}$ and $\chi^{(2)}$ be the two
low-energy stream functions. If the space were exactly invariant, each
composite $\chi^{(j)}\circ f$ would equal $\sum_k W_{kj}\,\chi^{(k)}$ up to an
affine function, for a $2\times2$ matrix $W$ representing the involution. We
fit $W$ by least squares, matching the Hessians of the two sides, which are
free of the affine gauge, at the quadrature nodes. The relative fit residual
is $6\times10^{-7}$ and the eigenvalues are $\pm1$ to within $10^{-7}$; the
corresponding eigenvectors give the approximately even and odd
representatives. The labels even and odd refer to
$\chi$; $f$ is a reflection, $\det R=-1$, so the forms \eqref{eq:streampot}
have the opposite parity. The two pentagon panels of
figure~\ref{fig:streampot} are normalized separately.

\section{Discussion}
\label{sec:discussion}

We have numerically constructed a Sasaki--Einstein metric on the irregular
link of the \(\dP2\) cone and approximated its two primitive harmonic basic
\((1,1)\)-forms, each of which gives a primitive imaginary self-dual
\((2,1)\)-form on the cone. The degree-\(14\) polynomial metric has a held-out
mean-squared Monge--Amp\`ere residual of \(6\times10^{-14}\).
The rational form ansatz enforces closedness, primitivity and
square-integrability; at \(N=16\), the larger of the two estimated relative
\(L^2\) distances to the harmonic representative of the fitted metric is
\(3.4\times10^{-5}\).

The metric calculations reproduce the canonical solution on \(\Tone\),
the GMSW curvature invariants on \(\Ypq\), and DHHKW's eigenvalues,
eigenfunction coefficient and pointwise Einstein defect on \(\dP3\); on
\(\dP2\) the integrated Riemann-squared curvature of the link agrees with the
conjectured closed form \cite{ArabiArdehali:2015iow,Nakayama:2026bls} to
\(2\times10^{-9}\).
The form calculation reproduces the HEK amplitude ratio on \(Y^{3,2}\)
to \(10^{-9}\) and yields a period-fixed basis on \(\dP3\).
The \(d-3\) cohomology directions are built into the admissible space, and
convergence is measured by the decrease of the \(d-3\) energies with the
enrichment order.

The negative controls record large residuals at obstructed Reeb vectors.
That is an observation, not a lower-bound theorem or a criterion for
non-existence.
Separately, orthonormalization removes the residual plateaus caused by the
poorly conditioned polynomial bases studied here.

Polynomial fits attain smaller residuals than the tested networks.
On \(\dP3\), the networks terminate at \(3.3\times10^{-10}\) with imposed
symmetry and \(1.16\times10^{-8}\) without it, while the polynomial reaches
the \(10^{-14}\) level and below. The networks nevertheless provide useful
checks of symmetry recovery and of residual behaviour in a second function
class. Both methods use the toric reduction, and extending a network to
non-toric geometry would require a different input representation.

Most polynomial fits take minutes or less on the CPU of an Apple M1 Max,
and the most expensive runs are the networks, at up to seventy-eight minutes;
appendix~\ref{app:robust} gives the costs, budgets and reproducibility of
every principal run.

The construction is restricted to toric cones of complex dimension three.
The polygon formulation applies to all Reeb orbit types, but our validation
covers only the targets studied. The remaining numerical limitations include
basis conditioning, finite sampling and the lack of external references
beyond \(\dP3\) and one integrated invariant on \(\dP2\). The harmonic-distance estimate controls the form error for
the fitted metric, not the error induced by replacing it with the exact
Sasaki--Einstein metric. Flux quantization, which fixes \(C\) in
\(G_3=C\,\omega_{2,1}\), and the warp-factor equation \eqref{eq:warp} are not
addressed here.

Two future directions use the metric and the forms constructed here. The
Kaluza--Klein spectroscopy of the $\dP2$ metric is reported separately; its
protected part, the short multiplets that contribute to the superconformal
index, is fixed independently of the metric by the Kohn--Rossi cohomology of
the link \cite{Eager:2012hx}, which supplies an exact check of that part of
the spectrum. The
second is the construction of metrics on smoothings of the cones. A
complex-structure deformation that gives the vanishing three-cycle a finite
size exists for the cone over $\dP2$, in one branch, and over $\dP3$, in two
branches of dimensions two and one, and not for
$\dP1$ \cite{Altmann:1994,Butti:2006deform,Franco:2005fd}, and it is on such
smoothed cones that a solution of
Klebanov--Strassler type would live \cite{Doran:2007zn}. Complete Calabi--Yau
metrics asymptotic to the cone exist on such smoothings
\cite{ConlonHein:2024,ConlonNghiem:2025}, the cone over $\dP2$ with its
irregular Reeb vector among the cases treated \cite{ConlonNghiem:2025}, but the
only one known in closed form is the deformed conifold \cite{Candelas:1989js}.
A smoothing breaks the
torus action, so on the deformed cone the polygon is gone and the problem
becomes a complex Monge--Amp\`ere equation for a K\"ahler potential. Such a construction requires methods beyond the two-variable toric
reduction used here.

\acknowledgments
We thank Richard Eager for pointing us to the curvature integral of
section~\ref{sec:dp2se}.
This work was supported by the National Research Foundation of Korea under the
grants RS-2025-00518906 and RS-2025-25457100. Part of this work was presented
at the SIMIS--APCTP meeting on AI methods in Theoretical Physics, held at the
Shanghai Institute for Mathematics and Interdisciplinary Sciences on
26--29 August 2026; N.K. thanks the organizers for their hospitality.
S.K. is supported by a KIAS Individual Grant (AP103401) via the Center for
Artificial Intelligence and Natural Sciences at Korea Institute for Advanced
Study.
Code implementing this method, and a first draft of this manuscript, were
developed with the assistance of Claude (Anthropic). Editorial revisions
were assisted by Codex (OpenAI). The problem formulation,
the validation design and the interpretation of the results are the authors' own, and the authors take full responsibility for the content.

\appendix

\section{The potential on the cone}
\label{app:potential}

For a cone potential with fixed Reeb vector $b$, the relation
$2G_{IJ}y^J=b_I$ gives
$\partial_I(y^JG_J-G)=b_I/2$. After fixing the additive constant in $G$,
this integrates to $y\cdot\nabla G-G=\ell_b/2$, and hence
\[
G(\lambda y)=\lambda G(y)+\tfrac12\lambda\ln\lambda\,\ell_b(y).
\]
In particular, $\Hess G$ is homogeneous of degree $-1$. Setting
$H\equiv G-\tfrac12\ell_b\ln\ell_b$ makes $H$ homogeneous of degree one
exactly, so $H=\ell_bF(y/\ell_b)$ and
$G=\tfrac12\ell_b\ln\ell_b+\ell_bF(t)$ with $F$ arbitrary on the slice. Our
$\Gcan$ is the choice of that $F$ as the Guillemin form on the slice,
$F_{\rm can}(t)=\tfrac12\sum_a\langle v_a,t\rangle\ln\langle v_a,t\rangle$,
and the closed form above and the implementation agree exactly, to the last
bit.

Restricting the cone equation to the slice makes its transverse form explicit
at every $b$. With $G=\tfrac12\ell_b\ln\ell_b+\ell_bF(s)$, $s=\mathsf{e}^{\mathrm{T}}y/\ell_b$
for $\mathsf{e}=(e_1,e_2)$ and $k\equiv \mathsf{e}^{\mathrm{T}}(1,0,0)^{\mathrm{T}}$, one finds
$\det\Hess_yG=\tfrac{|b|^2}{2\ell_b^{3}}\det\Hess_sF$ and
$2\partial_{y^1}G=3\ln\ell_b+3+6F-\nabla F\cdot(6s-2k)$, so on $\ell_b=1$ the
equation \eqref{eq:coneMA} reads
\begin{equation}
\ln\det\Hess_sF+6F-\nabla F\cdot(6s-2k)+\tilde c=0,
\qquad
\tilde c=c+3+\ln\tfrac{|b|^2}{2},
\label{eq:slicema}
\end{equation}
which is the transverse equation quoted in section~\ref{sec:setupgeom} with
$\Lambda=3$ and $\gamma=2k$; off the slice the $3\ln\ell_b$ cancels against the
$\ell_b^{-3}$ of the determinant, which is the radial invariance of the
residual. Nothing in this uses the rationality of $b$. The constants $c$ of
\eqref{eq:coneMA} and $\tilde c$ here differ by a number fixed by $b$ and are
not the same symbol. Both identities are checked to machine precision on the
$\dP3$ and $\dP2$ potentials.

MSY's canonical-plus-Reeb potential differs from $\Gcan$ by
$\tfrac12\,\ell_\infty\ln(\ell_\infty/\ell_b)$ with
$\ell_\infty=\langle\sum_av_a,y\rangle$, which is $\ell_b$ times a function
smooth on the closed polygon and is therefore absorbed into $\psi$.

\section{Robustness, convergence, and cost}
\label{app:robust}

All runs reported here were made on the CPU of an Apple M1 Max, in double
precision, with JAX 0.10.2 and SciPy 1.17.1; the evaluation limit of
$15{,}000$ quoted below and in section~\ref{sec:fingerprint} is SciPy's
default for \texttt{L-BFGS-B}. The reproducibility figures of
table~\ref{tab:reproduce} compare this build with a second one, JAX 0.4.30
and SciPy 1.13.1 on an Apple M4 Pro. The repository pins the remaining
versions.

We assess sensitivity to the sampling margin, initialization and numerical
implementation.
Shrinking the sampling margin from $2\times10^{-3}$ to $3\times10^{-5}$ moves
the held-out residual by a factor $1.4$ at degree $14$ and $1.8$ at degree
$18$, and retraining from a fresh
initialization moves the held-out residual by less than $5\%$. Positive
definiteness of the slice Hessian $\Hess G_P$ is checked on finite point sets
and reported as such: its smallest eigenvalue is $2.329$ on both $\dP3$ fits and
$1.711$ on the degree-$14$ $\dP2$ potential at $b^{\ast}$ that
section~\ref{sec:oneone} uses, the same to four digits over the training
samples (margin $2\times10^{-3}$) and over the $13{,}824$ and $11{,}520$ nodes
of the deterministic rule at $48$ nodes per direction, which approach the edges
to $1.5\times10^{-4}$ of the polygon's size.

Ill-conditioning of a truncated basis recurred three times beyond the
$\Ypq$ diagnosis of section~\ref{sec:whiten}, and the record is kept here. In
the $D_6$-invariant basis on $\dP3$ the averaged monomials are collinear enough
that their singular values span ten orders, and a relative threshold on them
discarded genuine directions without warning, at a cost of a factor $28$ in the
held-out residual at degree $14$ and $25$ at degree $18$; the remedy is to fix
the rank against the invariant count. In the harmonic forms of
section~\ref{sec:oneone} a plain tolerance cut on the form design silently
truncated the span, and a rank-revealing pivoted QR replaced it. And in
constructing the admissible space of section~\ref{sec:oneone-adm} the edge
constraints imposed on raw monomials of degree $16$--$20$ produced a null space
larger than the count allows, its spurious members having tangential second
derivatives $\chi_{zz}$ on the edges of $10^{-9}$ rather than zero; imposing the same constraints on an
$L^2$-orthonormalized basis removed them. In each case the symptom was the same,
a span whose dimension did not grow as the degree did, or grew past what a
count permits.

The distinction between iteration and evaluation limits is important for the
network fits. For the symmetry-free \(\dP3\) network, raising the production
iteration cap from \(6000\) to \(18{,}000\) reduces the residual from
\(4.82\times10^{-7}\) to \(4.18\times10^{-7}\), while a further increase to
\(54{,}000\) has no effect: both extended runs stop after \(13{,}739\)
iterations at the default \(15{,}000\) function-and-gradient evaluation limit.

Increasing the evaluation limit to \(150{,}000\) gives
\(2.38\times10^{-8}\) after \(137{,}697\) iterations, again at the limit.
With a limit of \(2\times10^6\), the run satisfies the convergence test after
\(698{,}344\) iterations and \(759{,}744\) evaluations, taking seventy-eight
minutes. Its held-out and in-sample residuals are
\(1.16\times10^{-8}\) and \(1.16\times10^{-9}\), respectively.
This is the endpoint of the reported run, not a lower bound for the network
class. The \(D_6\)-informed network meets its convergence test after \(7770\)
iterations and \(10{,}171\) evaluations at \(3.3\times10^{-10}\). At their
convergence endpoints the two networks therefore differ by a factor \(35\) in
held-out residual and \(75\) in evaluations, compared with a factor \(5\) in
residual at the production caps.

Widths, sample counts, caps and iterations for the principal runs are collected in
table~\ref{tab:cost}, and table~\ref{tab:reproduce} says how each reproduces.

\begin{table}[htbp]
\centering
\footnotesize
\begin{tabular}{llrrrl}
\toprule
step & ansatz & params & samples & L-BFGS cap & iterations used\\
\midrule
$Y^{2,1}$ & poly deg 8, ortho & 42 & 4096 & $2\times10^{4}$ & 453
(convergence)\\
$Y^{3,2}$ & poly deg 12, ortho & 88 & 4096 & $2\times10^{4}$ & 1052
(convergence)\\
sweep & poly deg 12--14, ortho & 88--117 & 4096 & $3\times10^{4}$ &
convergence\\
$\dP3$ & $D_6$-inv.\ deg 14 & 14 & 2048 & $2\times10^{4}$ & 161
(convergence)\\
$\dP3$ & $D_6$-inv.\ deg 18 & 21 & 8192 & $2\times10^{4}$ & convergence\\
$\dP2$ KE attempt & poly deg 6--16 & 25--150 & 3000 & $2\times10^{4}$ &
cap\\
$\dP3$ neural, $D_6$ & MLP, width 20 & 502 & 1024 & 3000 & cap\\
$\dP3$ neural, no $D_6$ & MLP, width 20 & 502 & 1024 & 6000 & cap\\
$\dP2$ neural & MLP, width 8--32 & 106--1186 & 3000 & $10^{6}$ & cap, at
$10^{5}$ evals\\
$\Tone$ recovery & MLP, width 16 & 338 & 1024 & --- & Adam, 1500\\
\bottomrule
\end{tabular}
\caption{Widths, sample counts, caps and iterations used for the principal runs
reported in this paper; the $\dP2$ polynomial at $b^{\ast}$
(table~\ref{tab:twoclasses}), the released network runs of this appendix and
the harmonic-form computations of section~\ref{sec:oneone} are described where
they appear. The $\Tone$ recovery run is the gate of section~\ref{sec:t11},
posed in the transverse form. Which budget binds an entry reading ``cap''
differs by row, and the difference decides what raising a cap is worth. The
symmetry-free $\dP3$ network stops on \texttt{scipy}'s default limit of
$15{,}000$ function and gradient evaluations rather than on the iteration cap
beside it, which is why raising that cap alone changed nothing there, and
appendix~\ref{app:robust} releases the evaluation limit to $2\times10^{6}$
instead. The $D_6$-informed network is the other way round: it reaches $3000$
iterations after some $3900$ evaluations, so the iteration cap is what stops
it, and releasing that alone carries it to its convergence test at
$3.3\times10^{-10}$. The $\dP2$ neural row already has the evaluation limit
released to $10^{5}$.}
\label{tab:cost}
\end{table}

\begin{table}[htbp]
\centering
\footnotesize
\begin{tabular}{lll}
\toprule
exit condition & across builds & quoting rule\\
\midrule
convergence test & ratio $0.99$--$1.00$ & the digits quoted reproduce\\
iteration cap & fourth significant figure moves & four figures at most\\
stored minimizer, one draw & one-sigma scatter $7\%$ & large-sample value\\
\bottomrule
\end{tabular}
\caption{How a run reproduces across software builds, by the condition on which it exits, and the quoting rule each case earns.}
\label{tab:reproduce}
\end{table}

Table~\ref{tab:twoclasses} compares costs on the \(\dP2\) target at
\(b^{\ast}\), on the same machine. A degree-\(16\) polynomial
evaluation takes \(26\) ms and a network evaluation \(10\) ms on \(3000\)
samples. The main difference is the number of evaluations: the polynomial
converges after about \(2.3\times10^3\), while the network exhausts
\(10^5\) evaluations with a residual five orders of magnitude larger.

The capped pentagon run does not determine its asymptotic residual. On $\dP3$, where the symmetry-free network was carried
to its convergence test, the trajectory recorded every $10{,}000$ iterations
falls as iterations$^{-0.38}$ over its last sixty-six samples, with a
correlation of only $0.83$ --- a slow approach to a floor rather than a power
law, and the run terminates at $1.16\times10^{-8}$. On the pentagon at $b^{\ast}$
the network is still capped at $4.1\times10^{-10}$, so its floor there is not
known, and we make no extrapolation for it.

\begin{table}[htbp]
\centering
\footnotesize
\begin{tabular}{lcccccc}
\toprule
ansatz & params & evaluations & iterations & wall clock & exit & held-out\\
\midrule
polynomial, deg $12$ & $88$ & $643$ & $617$ & $17$\,s & convergence &
$1.7\times10^{-12}$\\
polynomial, deg $14$ & $117$ & $1013$ & $969$ & $25$\,s & convergence &
$6.0\times10^{-14}$\\
polynomial, deg $16$ & $150$ & $2307$ & $2131$ & $60$\,s & convergence &
$2.5\times10^{-15}$\\
network, width $16$ & $338$ & $100{,}001$ & $90{,}919$ & $994$\,s & cap &
$6.0\times10^{-10}$\\
\bottomrule
\end{tabular}
\caption{What the two function classes cost on the $\dP2$ pentagon at
$b^{\ast}$, on the same machine with nothing else running. The
network is given the released evaluation budget of section~\ref{sec:fingerprint};
the polynomial is given the same limit and does not approach it. The network
entry is a separate run from the width sequence of that section, whose
width-$16$ point at $b^{\ast}$ lands at $4.4\times10^{-10}$; the two differ by
the sample draw and not by the budget. The exit
column is the substance of the table: the polynomial stops because it has
converged, the network stops because it has run out of budget, and the two
things are not comparable as accuracies.}
\label{tab:twoclasses}
\end{table}

The form matrices of section~\ref{sec:oneone-energy} are checked in the $L^2$-orthonormalized basis at $N=16$: $T$ has three
eigenvalues at $-1.0000$ on the hexagon and two on the pentagon; the remaining
eigenvalues have absolute values below $5\times10^{-9}$ and $8\times10^{-7}$,
respectively. The identity $S=\tfrac12(M+T)$ holds to
$2\times10^{-8}$ and $1\times10^{-5}$ in the largest entry; at $N=20$ the
conditioning of the enriched basis limits these matrix statements to
$10^{-5}$. The low-energy vectors have tangential second derivatives
$\chi_{zz}$ on the edges at roundoff, $10^{-15}$, and re-integrating them, held
fixed, at $24$ to $128$ nodes per direction reproduces numerator and
denominator to every digit printed (table~\ref{tab:quadscan}). Quadrature stability does not bound the arithmetic
of the basis change, so each fixed stream function is also scored without it,
by integrating the density $\tfrac12\det\Hess G_P\,(u^{ij}\chi_{ij})^2$ of
section~\ref{sec:oneone-energy} directly from its coefficients; this
reproduces every printed digit, and returns the two $A_2$ energies equal to
ten digits, consistent with the exact symmetry degeneracy. Two algebraically
equivalent routes to the type residual, and extended-precision accumulation of
the quadrature sums, move these energies by at most $1.3\times10^{-12}$ and
$4.5\times10^{-13}$ relative, over the five vectors at $N=20$ on the hexagon
and $N=16$ on the pentagon. The eigenvalues of the orthonormalized problem
differ from the fixed-vector values by $5\times10^{-5}$ relative at $N=20$ on
the hexagon, where they split the exact pair by $1.5\times10^{-4}$, and by
$4\times10^{-7}$ and $3\times10^{-8}$ at $N=16$ on the pentagon; the energies
tabulated in this paper are the fixed-vector values. All of this is
consistency of those energies, not a bound on the errors made in computing the
vectors, and the $Y^{3,2}$ and conifold numerators, $4\times10^{-18}$ and
$10^{-29}$, are floors of the arithmetic rather than measurements.

\begin{table}[htbp]
\centering
\footnotesize
\begin{tabular}{lc}
\toprule
low-energy vector & $\|\theta_+\|^2$\\
\midrule
hexagon, $N=16$, $\theta_{1,2}$ & $1.171350\times10^{-9}$\\
hexagon, $N=16$, $\theta_{3}$ & $5.562586\times10^{-9}$\\
hexagon, $N=20$, $\theta_{1,2}$ & $2.454048\times10^{-12}$\\
hexagon, $N=20$, $\theta_{3}$ & $1.236379\times10^{-11}$\\
pentagon, $N=16$, $\theta_1$ & $1.386958\times10^{-10}$\\
pentagon, $N=16$, $\theta_2$ & $5.722118\times10^{-10}$\\
\bottomrule
\end{tabular}
\caption{Quadrature stability of the low-energy vectors of the admissible space. Each vector is fixed at $48$ nodes per direction and re-integrated, numerator and denominator separately, at $24$, $48$, $96$ and $128$ nodes. The numerator is the entry shown at all four orders, to the seven digits printed, and the denominator is $1.0000000$ at all four; only the numerator is listed because nothing else varies. The entries are absolute, the denominators being normalized to one at $48$.}
\label{tab:quadscan}
\end{table}

\section{The \texorpdfstring{$Y^{p,q}$}{Y(p,q)} metric}
\label{app:ypq}

Sections~\ref{sec:sweep} and~\ref{sec:oneone-closed} refer to the coordinate
$y$ of the GMSW metric, and this appendix records what is needed of it. Its
symbols are theirs, so that the equations can be checked against the source as
printed: $\psi$, $\theta$, $\phi$ and $\alpha$ are their angles, $a$ and $c$
their constants, and $y$ a single coordinate rather than the cone coordinates
$y^I$, none of them the objects those letters denote elsewhere. The angles and
the constants occur only here; $y$ and its endpoints $y_1,y_2$ are used in the
main text as well, and there too they are GMSW's. In the
form of \cite[(2.1)--(2.3)]{Gauntlett:2004yd}, with their constant $c$ set to
one,
\begin{equation}
\begin{aligned}
ds^2_{Y^{p,q}}
&=\frac{1-y}{6}\,(d\theta^2+\sin^2\theta\,d\phi^2)
+\frac{dy^2}{w(y)\,q(y)}
+\frac{q(y)}{9}\,(d\psi-\cos\theta\,d\phi)^2\\
&\quad+w(y)\Bigl[d\alpha+\frac{a-2y+y^2}{6(a-y^2)}\,(d\psi-\cos\theta\,d\phi)\Bigr]^2 ,
\end{aligned}
\label{eq:gmswmetric}
\end{equation}
\begin{equation}
w(y)=\frac{2(a-y^2)}{1-y},
\qquad
q(y)=\frac{a-3y^2+2y^3}{a-y^2},
\label{eq:gmswwq}
\end{equation}
which locally satisfies $\Ric=4g$ wherever the metric is nondegenerate;
global regularity restricts the parameters and angular identifications. The coordinate $y$ runs over
$[y_1,y_2]$, where $y_1<0<y_2$ are the two smaller roots of the cubic
$a-3y^2+2y^3=0$, and the Reeb vector is a combination of $\partial_\psi$ and
$\partial_\alpha$ whose orbits close exactly when $4p^2-3q^2$ is a perfect
square \cite[(3.8)]{Gauntlett:2004yd}. The integers $(p,q)$ enter through the
separation of the two roots,
\begin{equation}
\delta_y\equiv y_2-y_1=\frac{3q}{2p},
\qquad
y_1=\tfrac12\Bigl(1-\delta_y-\sqrt{1-\delta_y^2/3}\Bigr),
\qquad
a=3y_1^2-2y_1^3 ,
\label{eq:gmswroots}
\end{equation}
which is the $\delta_y$ of table~\ref{tab:sweep}. On $Y^{3,2}$, $\delta_y=1$ gives
$y_1=-0.408248$, $y_2=0.591752$ and hence the ratio $11.898979$ quoted in
section~\ref{sec:oneone-closed}. The coordinate $y$ is a moment map for one
circle in the torus, so in the symplectic frame of section~\ref{sec:setupgeom}
it is an affine function of the coordinates $y^I$ of the moment cone restricted
to the link; its normalization, in which the $S^2$ factor carries $(1-y)/6$,
is the one in which the HEK form has eigenvalues
$\pm(1-y)^{-2}$ \cite{Herzog:2004tr}.

\section{Conventions of the comparison with DHHKW}
\label{app:conventions}

For $\theta=i\partial\bar\partial\mu$ in the symplectic frame,
\[
A_{ij}=\tfrac12\partial_i(u^{jk}\partial_k\mu),
\qquad
\mathrm{tr}A=\tfrac12\triangle\mu,
\]
where $u=(\Hess G_P)^{-1}$ and $\triangle$ is the slice Laplacian
of section~\ref{sec:graders} \cite[(3.46)]{Doran:2007zn}.
Table~\ref{tab:conventions} gives the conversions used in
sections~\ref{sec:dp3grading} and~\ref{sec:oneone-dp3}.

\begin{table}[htbp]
\centering
\begin{tabular}{ll}
\toprule
quantity & conversion\\
\midrule
coordinates & $\tilde s=3s$\\
symplectic potential & $G_{\rm DHHKW}(\tilde s)=3G_P(\tilde s/3)+\text{affine}$\\
metric & $g_{\rm DHHKW}=3\tilde g_T$\\
Laplacian & $\triangle_{\rm DHHKW}=\tfrac13\triangle$\\
form potential & $\mu(s)=-2\mu^{\rm src}(3s)$ (calibrated)\\
\bottomrule
\end{tabular}
\caption{Conversions from DHHKW's conventions to the slice conventions of this paper.}
\label{tab:conventions}
\end{table}

The factor $-2$ is calibrated by DHHKW's contraction $\tfrac23$ for
$\theta_2$ \cite[(6.14)--(6.15)]{Doran:2007zn}; in their metric,
the contraction is $\tfrac16\triangle\mu=-\tfrac13\triangle\mu^{\rm src}$,
with the source potential expressed in $s$.
Their value $2$ for $\omega=\tfrac12\sum_a\theta_a$ follows from the same
anchor by dihedral symmetry. An independent check uses the slice
K\"ahler potential
\[
\mu_T=2(s\cdot\nabla G_P-G_P),\qquad \triangle\mu_T=4.
\]
DHHKW's printed potential for $\omega$ agrees with $-\tfrac32\mu_T$ up to
a constant to $2.0\times10^{-4}$ rms over the hexagon. The conversion then
gives $\omega=3\omega_T$, as required by $g_{\rm DHHKW}=3\tilde g_T$.

\section{Admissibility and completeness of the trial space}
\label{app:admissibility}

For a compact smooth toric Sasakian link $Y$ with slice polygon $P$, let
$\mathcal A(P)$ denote the smooth functions on $\overline P$ whose edge
traces are affine. Smoothness means extendibility to a neighbourhood of
$\overline P$. The normalized moment map identifies $P$ with the quotient of $Y$
by the full torus for every Reeb vector; a torus-invariant
function on $Y$ is smooth exactly when it is a smooth function of the moment
coordinates, so smoothness on $\overline P$ is a statement about $Y$ and does
not involve the leaf space of the Reeb flow. We justify the claims used in
section~\ref{sec:oneone}:
$\theta_\chi$ is smooth for $\chi\in\mathcal A(P)$, its primitive basic class
is determined by vertex values modulo affine functions, and each such class
has a unique harmonic stream function modulo that gauge; the trial space
\eqref{eq:admspace} is dense in $\mathcal A(P)$ and the Gram matrix $T$ of
the intersection pairing has rank $d-3$. The argument uses one elementary
computation, the behaviour of forms near a collapsing circle in polar
coordinates; one application of Stokes' theorem; and two results taken from
the literature, the basic Hodge theorem, which gives every basic class a
unique harmonic representative
\cite{ElKacimiAlaouiHector:1986,ElKacimiAlaoui:1990}, and the vanishing of
the off-diagonal basic Hodge numbers of a toric Sasakian manifold, which
makes every degree-two basic class of type $(1,1)$
\cite[Cor.~8.4]{GoertschesNozawaToben:2016}. Neither an Einstein metric nor a
rational Reeb vector is required for these statements. We then show that a
globally polynomial stream function with affine edge traces represents no
nontrivial class on the polygons used here, the square excepted.

\emph{Forms near a collapsing circle.}
Near an interior point of edge $a$, the link is locally a product of 
the surviving two-torus, an interval along the edge with
coordinate $z$, and a plane. The circle generated by $v_a$ acts by rotation on 
the plane, and collapses at the origin. 
Let $(\rho,\vartheta)$ be polar coordinates on the plane; the
moment coordinate is $x=\ell_a$, a smooth positive multiple of $\rho^2$.
Rotation-invariant smooth functions on the plane are smooth functions of
$x$, and rotation-invariant smooth one-forms are
$p(x)\,dx+r(x)\,x\,d\vartheta$ with $p,r$ smooth, since
$x\,d\vartheta\propto\rho^2d\vartheta=X\,dY-Y\,dX$ in Cartesian coordinates
$(X,Y)$; $d\vartheta$ itself is not smooth at the origin, while
$dx\wedge d\vartheta\propto dX\wedge dY$ is. Set $n_{a,j}=\langle v_a,e_j\rangle$.
In a lattice basis containing $v_a$, the difference of $\nu_j$ and
$n_{a,j}\,d\vartheta$ is a smooth angular form of the surviving two-torus, so
for a coefficient vector $\mathbf a$ smooth on $\overline P$,
\[
 \mathbf a\cdot\nu=(n_a\cdot\mathbf a)\,d\vartheta+(\text{smooth}),
 \qquad
 d(\mathbf a\cdot\nu)=d(n_a\cdot\mathbf a)\wedge d\vartheta+(\text{smooth}).
\]
Writing $n_a\cdot\mathbf a=k_a(z)+x\,g_a(x,z)$, the one-form is smooth
precisely when $k_a=0$, and its exterior derivative,
\[
 d(k_a+xg_a)\wedge d\vartheta
 =k_a'(z)\,dz\wedge d\vartheta+g_a\,dx\wedge d\vartheta+dg_a\wedge(x\,d\vartheta),
\]
is smooth precisely when $k_a$ is constant along the edge, the last two
terms being smooth. Near a vertex two circles collapse, the local model is
$\mathbb C^2\times S^1$ with moment coordinates proportional to
$\rho_1^2,\rho_2^2$, and the two conditions hold simultaneously.

For $\mathbf a_\chi=(\chi_2,-\chi_1)$, the vector $(-n_{a,2},n_{a,1})$ is
tangent to edge $a$, so $n_a\cdot\mathbf a_\chi=(-n_{a,2},n_{a,1})\cdot\nabla\chi$
is the tangential derivative of $\chi$ along the edge; its constancy is the
affine edge-trace condition, and its vanishing says that $\chi$ is constant
along the edge. Thus every $\chi\in\mathcal A(P)$ gives a smooth basic
two-form $\theta_\chi$, closed and primitive by construction, and if a smooth
$f$ vanishes on $\partial P$, then $\alpha_f=\mathbf a_f\cdot\nu$ is a smooth
basic one-form and $\theta_f=d_B\alpha_f$.

\emph{Classes and vertex values.}
Conversely, if $\theta_\chi=d_B\beta$, average the smooth basic one-form
$\beta$ over the torus, which leaves $d_B\beta=\theta_\chi$ unchanged,
$\theta_\chi$ being invariant, and write
$\beta=p_i\,ds_i+(\mathbf a_\beta)_j\nu_j$ on the dense orbit.
The angular coefficients are $\beta(X_{e_j})$, where
$X_{e_j}=(e_j)_I\partial_{\phi_I}$ is the smooth torus vector field dual to
$\nu_j$, $\nu_k(X_{e_j})=\delta_{kj}$; they are smooth invariant functions,
hence smooth on $\overline P$. Comparing mixed components gives
$\mathbf a_\beta=\mathbf a_\chi-k$ for a constant vector $k$.
Smoothness of $\beta$ forces $n_a\cdot\mathbf a_\beta=0$ on each edge.
Choose an affine $\ell$ with $\mathbf a_\ell=k$; then $\chi-\ell$ is
constant along every edge. Continuity makes these constants equal, so,
after shifting $\ell$, $\chi-\ell$ vanishes on the boundary. In the other
direction, if
the vertex values of $\chi,\chi'\in\mathcal A(P)$ differ by an affine $\ell$,
then $\chi-\chi'-\ell$ has affine edge traces vanishing at the vertices,
hence vanishes on $\partial P$, and $\theta_\chi-\theta_{\chi'}$ is basic-exact by
the preceding paragraph. Thus two stream functions represent the same basic
class exactly when their vertex values differ by affine data.

\emph{The harmonic representative.}
By the basic Hodge theorem, every primitive basic class has a unique
harmonic representative $\theta_{\mathrm{harm}}$, and by the vanishing of
the off-diagonal basic Hodge numbers it is a real form of type $(1,1)$.
Wedge multiplication by $\omega_T$ commutes with the basic Laplacian, so
$\theta_{\mathrm{harm}}\wedge\omega_T$ is a harmonic basic four-form, a
constant multiple of $\omega_T\wedge\omega_T$, and the constant vanishes
because the class is primitive; a real primitive $(1,1)$-form is
anti-self-dual. The torus acts by isometries commuting with $\xi$ and
trivially on basic cohomology, so uniqueness makes $\theta_{\mathrm{harm}}$
torus invariant. As in section~\ref{sec:oneone-reduce}, closedness, boundary
smoothness and anti-self-duality then leave only the mixed components,
$\theta_{\mathrm{harm}}=A_{ij}\,ds_i\wedge\nu_j$, and the coefficients
$A_{ij}$ are smooth on $\overline P$: $\iota_{X_{e_j}}\theta_{\mathrm{harm}}=-A_{ij}ds_i$
is a smooth invariant one-form without angular components, and by the local
model above its coefficients are smooth functions of the moment coordinates,
at the edges and at the corners. The two integrations of
section~\ref{sec:oneone-reduce}, from $A$ to $\mathbf a$ and from
$\mathbf a$ to $\chi$, are integrals along straight segments in the convex
polygon $\overline P$ and preserve smoothness up to the boundary, so
$\theta_{\mathrm{harm}}=\theta_\chi$ for a $\chi$ smooth on $\overline P$;
smoothness of $\theta_{\mathrm{harm}}$ places $\chi$ in $\mathcal A(P)$, and
type $(1,1)$ gives \eqref{eq:oneoneeq}.

Every vertex vector $c$ has a Wachspress extension $\sum_ac_a\varphi_a$.
The harmonic representative of its class has a stream function by the
preceding construction; the exactness criterion allows an affine adjustment
to recover precisely $c$. Uniqueness of the harmonic representative makes
two such solutions differ only by an affine function, and an affine function
vanishing at the $d$ vertices is zero. Hence the primitive basic classes are parametrized by
$d$ vertex values modulo three affine directions, giving dimension $d-3$
without assuming a cohomological count.

\emph{Density and the intersection pairing.}
For the density of \eqref{eq:admspace}, let $\chi\in\mathcal A(P)$ have
vertex values $c_a$; then $\chi-\sum_ac_a\varphi_a$ has affine edge traces
vanishing at the vertices and so vanishes on $\partial P$. Smooth division at
edges and corners writes any smooth function vanishing on $\partial P$ as
$\prod_a\ell_a\cdot q$ with $q\in C^\infty(\overline P)$, so $\chi$ has the
form \eqref{eq:admspace} with a smooth $q$ in place of a polynomial;
polynomial approximation of $q$ in $C^2$ gives convergence of the stream
functions in $C^2$. The polar-coordinate estimates above give, for the fixed
metric,
\[
 \bigl\|\theta_{\prod_a\ell_a\,(q-q_N)}\bigr\|_{L^2}
 \le C_g\|q-q_N\|_{C^2(\overline P)}\longrightarrow0.
\]
This proves the required density of \eqref{eq:admspace}.
For closed basic $\beta$, Stokes' theorem gives
\[
 \int_Y\eta\wedge d_B\alpha\wedge\beta
 =\int_Yd\eta\wedge\alpha\wedge\beta=0,
\]
because the final integrand has transverse degree five. The intersection
pairing therefore depends only on the classes. For primitive basic classes,
anti-self-duality of their harmonic representatives gives
\[
 \int_Y\eta\wedge\theta\wedge\theta'
 =-\langle\theta_{\mathrm{harm}},\theta'_{\mathrm{harm}}\rangle_{L^2(Y)},
\]
with the same transverse metric and volume convention on both sides.
For $\theta'=\theta$ this is $-\|\theta_{\mathrm{harm}}\|_{L^2(Y)}^2$.
Consequently $T$ has rank $d-3$, is negative definite on the
vertex directions, and has the bubble directions as its kernel.

\emph{Polynomial stream functions.}
A polynomial whose restriction to an edge is affine is affine on the whole
line through it, so a globally polynomial stream function with affine edge
traces must have traces that agree where non-parallel edge lines intersect,
also outside $P$. Let $L_a(c;s)$ be the affine trace on the line $\ell_a=0$
determined by its two endpoint values in $c$, and let $p_{ab}$ be the
intersection of two non-parallel edge lines; define
\[
 (C_Pc)_{ab}=L_a(c;p_{ab})-L_b(c;p_{ab}).
\]
Affine vertex data lie in $\ker C_P$, and on the hexagon, pentagon and
$Y^{3,2}$ quadrilateral exact row reduction of $C_P$ at the Reeb vectors
used here gives rank $d-3$, so $\ker C_P$ is exactly the affine vertex data.
Subtracting the corresponding affine function leaves a polynomial vanishing
on every edge line, hence divisible by $\prod_a\ell_a$: only affine plus
bubble functions remain, whose nonzero forms have $\mathcal E=\tfrac12$ by
the pairing above, and no nontrivial class is represented. The square, whose
opposite edges are parallel, is the exception: it admits the quadratic
$\chi=uv$ in edge-aligned affine coordinates, which carries its one
nontrivial class.

\bibliographystyle{JHEP}
\bibliography{refs}

\end{document}